\documentstyle[psfig]{mn}

\newcommand{\ctabla}[2]{\makebox[7mm][r]{#1}$\pm$\makebox[5mm][l]{#2}}
\newcommand{\rtabla}[1]{\makebox[5mm][r]{$\sim$}\makebox[7mm][r]{#1}}

\title[Stellar indices and kinematics in Seyfert 1]
  {Stellar indices and kinematics in Seyfert 1 nuclei}

\author[L. Jim\'enez--Benito et al.]
  {Luis Jim\'enez--Benito,$^1$ Angeles I.~D\'{\i}az,$^1$ Roberto Terlevich$^2$
\thanks{Visiting Professor, INAOE, Puebla,
Mexico} and Elena Terlevich$^3$
\thanks{Visiting Fellow, IoA, Cambridge}\\
$^1$Departamento de F\'{\i}sica Te\'orica C-XI, Universidad Aut\'onoma de 
Madrid, Cantoblanco, 28049, Madrid, Spain\\
$^2$Institute of Astronomy, Madingley Road, Cambridge, CB3 0HA, United 
Kingdom\\
$^3$Instituto Nacional de Astronom\'{\i}a, \'Optica y Electr\'onica, 
Tonantzintla, Puebla, Mexico}

\pagerange{\pageref{firstpage}--\pageref{lastpage}}
\volume{000}
\pubyear{1999}

\begin{document}

\label{firstpage}

\maketitle


\begin{abstract}
We present spectra of 6 type 1 Seyfert galaxies, 2 Seyfert 2,
a starburst galaxy and a compact narrow line radiogalaxy, taken
in two spectral ranges centered around the near--IR \hbox{Ca\,{\sc ii}} triplet
($\sim$ 8600\AA\ ), and the  Mgb stellar feature at 5180\AA .
We measured the equivalent width (EWs) of these features and the Fe$_{52}$
and Fe$_{53}$ spectral indices.

We found that the strength of the IR \hbox{Ca\,{\sc ii}} triplet (CaT)
in type 1 Seyfert galaxies with prominent central point sources, is larger
than what would be expected from 
the observed strength of 
the blue indices. 
This could be explained by the presence of red supergiants
in the nuclei of Seyfert 1 galaxies. On the other hand, the blue indices of these galaxies
could also be diluted  by the strong \hbox{Fe\,{\sc ii}} multiplets that can be seen in
their spectra.

We have also measured the stellar and gas velocity dispersions of the galaxies in
the sample. The stellar velocity dispersions were measured using both, the Mgb and 
CaT stellar features. The velocity dispersion
of the gas in the narrow line region (NLR) was measured using the strong emission lines 
\hbox{[O\,{\sc iii}]} $\lambda \lambda$5007, 4959 and \hbox{[S\,{\sc iii}]} 
$\lambda$9069. We compare the gas and star velocity dispersions 
and find that both magnitudes
are correlated in Seyfert galaxies. 

Most of the Seyfert 1 we observe have stellar velocity 
dispersion somehow greater than that of the gas in the NLR. 

\end{abstract}




\section{Introduction}

The presence of young massive stars in the nuclear regions of Seyfert galaxies was
strongly suggested by the detection of the near--IR
absorption \hbox{Ca\,{\sc ii}} triplet 
($\sim$ 8600\AA ) (CaT) in a sample of active galactic nuclei by 
Terlevich, D\'{\i}az \& Terlevich \shortcite{TDT90} (hereafter TDT90). This stellar feature 
depends strongly on gravity
and only weakly on metallicity, and is known to be specially strong in
young red supergiants (Jones, Alloin \& Jones 1984; D\'{\i}az, Terlevich 
\& Terlevich 1989).  
The analysis of this feature led TDT90 to conclude that strong 
starbursts should be present in the nuclei of  
Seyferts 2, since, despite the weakness or dilution observed in the blue stellar absorption lines,  
the CaT was found to be very strong in the nuclear spectra of 
the twelve galaxies of this type that were observed. Moreover, the only three Seyfert 1 
galaxies included in their sample also show CaT in absorption, 
suggesting at least some contribution by a young stellar population.

Subsequent IR spectroscopy 
(1.5 to 2.3 $\mu$m) of normal and active galaxies 
performed by Oliva et al. (1995) allowed them to conclude that the Seyfert 2 nuclei of their 
sample were compatible with evolution from a pre-cursor starburst.
They also find that the 1.6 to 2.3 $\mu$m stellar continuum of Seyfert nuclei is 
too red to be accounted for by a non-thermal continuum but is compatible with reprocessed 
radiation from hot dust.

Cid Fernandes \& Terlevich (1992;1993;1995), in a critical 
analysis of the simple unified scenario \cite{Ant93}, 
proposed that the observed strong CaT in Seyfert type
2 galaxies with strong and blue optical continuum plus the absence of broad 
lines combined with the low continuum polarization, were the result of
the presence in the nuclear region of unpolarized starlight from very young 
stars, i.e.~a nuclear/circumnuclear starburst or star forming toroid.  
This simple suggestion seems 
able to overcome most of  the  difficulties  faced  by  the  basic
unified model for Seyfert 2, while  preserving  its  attractive
features.

Schmitt, Storchi--Bergmann \& Cid Fernandes \shortcite{SSC98}
have recently found that the spectra of many Seyfert 2 can be modelled by the
sum of the spectra of a young stellar cluster (age $\sim$ 100 Myr) and an old 
one (age $\sim$ 10 Gyr), and that these
models reproduce the observations better than the traditional ones
consisting of a blue featureless continuum (BFC) and an old stellar population,
thus confirming Cid Fernandes \& Terlevich suggestions.

The question about the origin of the nuclear continuum in Seyfert 2, 
has given rise to searches of young stars in their nuclei. 
Heckman et~al. \shortcite{Heck97} and
G\'onzalez Delgado et~al. \shortcite{Gon98} have presented
high resolution UV images, taken with the HST, of 4  Seyfert 2 galaxies, finding
compact nuclear starbursts in all of them. They have also found spectral 
features from young hot stars in the UV spectra of these 4 galaxies. 
Their main conclusion is
that, in all the galaxies they have studied, the observed continuum is 
exclusively due to a nuclear/circumnuclear young starburst and
the energy emitted by the nuclear starburst is, at least, of the same order
as the one produced by the buried AGN.
Powerful starbursts have also been found in the nuclei
of several LINERS by Colina et~al. \shortcite{Col98} and Maoz et~al. 
\shortcite{Maoz98} and very recently in a QSO by Brotherton et~al. \shortcite{BBS99}.
 
If the results presented above can be generalized to all type 2 Seyferts 
then, we could conclude that nuclear starbursts 
ought to play an important
role in the total energy emitted in their nuclear regions.
Furthermore, since in unified models Seyferts 1 and 2 are not 
physically different kinds of object, but the consequence of a different 
viewing angle, one can conclude that, if nuclear 
starbursts are found to be energetically important in
Seyfert 2, so they should  be in Seyfert 1. 

However, the starbursts in Seyfert 1 nuclei, if present, may be somewhat different
from those in Seyfert 2: the narrow H$\alpha$ + \hbox{[N\,{\sc ii}]} emission in 
Seyfert 2 galaxies is more extended than in Seyfert 1 
hosts (Pogge 1989; Gonz\'alez Delgado \& P\'erez 1993).
Also, the galaxies with type 2 Seyfert nuclei have 
enhanced infrared emission from their disks compared to those with type 1 
nuclei or with normal spiral galaxies \cite{Mai95}. 
From their near--IR observations Oliva et~al. \shortcite{Oli95} found that the $M/L$ ratio  
in Seyfert 1 is similar to the one shown by normal early type spiral galaxies, 
whereas this ratio is found to be lower in Seyfert 2, pointing to  younger stellar populations.
Also, Gonz\'alez Delgado et~al. \shortcite{Gon97}
found that circumnuclear star-forming rings are more common
in Seyfert 2 than in Seyfert 1. Moreover the few UV HST images that exist 
of type 1 Seyferts do not
show compact starbursts in their nuclei but point sources.

In this paper we use the fact that the nuclear light output in Seyfert 1
galaxies has little contamination from the surrounding bulge to search for 
signatures of a nuclear starburst. If,  for example, strong CaT absorptions 
were detected in the unresolved nuclear component, because of the limited 
surface brightness, they could not be due to bulge contamination. A starburst 
signature should be possible to detect, even in the presence of a dominant 
nuclear component. In other words, if the result from 
Heckman et~al. \shortcite{Heck97} and Gonz\'alez Delgado et~al. \shortcite{Gon98}
can be generalized -- i. e. that the luminosity of 
the starburst is equal or larger than that of the AGN -- then the surface 
brightness of the stellar population giving rise to the CaT feature should 
be much higher than that from the old bulge component in order to be 
detected on top of the dominant AGN spectrum and therefore detectable in 
the presence of the nuclear continuum.

We have looked for signatures of young stars in the optical/NIR spectra of 6 type 1 Seyfert 
galaxies. Although, in principle, the UV would be the best spectral band  to detect absorptions 
from young stars, the strong broad line region (BLR) contamination at the 
wavelengths of the stronger stellar features 
(\hbox{[C\,{\sc iv}]} $\lambda$1550 \AA , \hbox{[Si\,{\sc iv}]} 
$\lambda$1400 \AA ) makes this method not viable.
On the other hand, as discussed by TDT90,
the region around the near--IR CaT is relatively free of strong emissions.
\begin{footnote}{The \hbox{O\,{\sc i}} $\lambda$8400 \AA\ and the  
\hbox{[Fe\,{\sc II}]} $\lambda$8617 \AA\ lines are 
exceptions to this. The first 
line is broad and prominent in luminous Seyfert 1 galaxies, while the second one 
is narrow and seems to be present in some starburst and Seyfert 2 galaxies.}
\end{footnote}

It is known that the stellar kinematics in the nuclei of Seyfert galaxies is 
similar to that of normal spiral galaxies, since both kinds follow the 
same Faber--Jackson relation (TDT90; Nelson \& Whittle 1996). 
It is also known that the gas motions in the 
narrow line region (NLR) of Seyfert galaxies seem to be dominated by the gravitational field
of the bulge, since the width of the \hbox{[O\,{\sc iii}]} $\lambda$ 5007 
emission line is correlated
with the nuclear stellar velocity dispersion (Wilson \& Heckman 1985; 
TDT90; Nelson \& Whittle 1996).
TDT90 found a small population of type 2 Seyfert 
galaxies with gas velocities much larger than the stellar ones. The existence
of this population was confirmed by Nelson \& Whittle \shortcite{NW96}. 

The kinematics of the stars and gas in the nuclei of our sample galaxies was
also studied in this paper, improving previous results by adding 6 new Seyfert 1 galaxies
to the above refered samples.

\begin{figure}
\centerline{
\psfig{figure=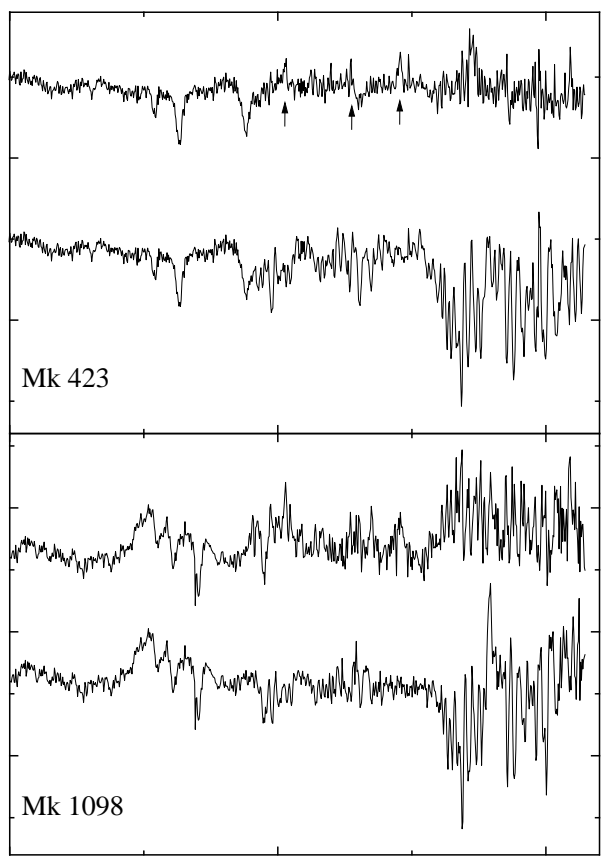,width=8cm,clip=}}
\caption{Elimination of atmospheric absorption bands. In both frames, the top
spectrum is the atmospheric absorption corrected one.
In Mk~1098 (bottom), the correction method caused 
broad ``emission bands'' to appear. On the contrary in
Mk~423 (top), only weak and narrow ``emission lines'' (marked 
with arrows) appear after the correction procedure.}
\label{comp}
\end{figure}


\begin{table}
\caption{\label{obs}Observational parameters.}

\begin{tabular}{lcc}
\hline
$\lambda_{c}$ & 5100 \AA & 8700 \AA \\
\hline
Dates & 3--4 May 1991 & 3--4 May 1991 \\
Telescope & WHT & WHT \\
Spectrograph & ISIS & ISIS \\
Grating & 600B & 316R \\
Detector & CCD TEK 1 & CCD TEK 2 \\
Filter & --- & CG495 \\
Spectral range & 4706--5607\ \AA & 7983--9573 \AA \\
Dispersion & 0.73 \AA\ pixel$^{-1}$ & 1.39 \AA\ pixel$^{-1}$ \\
Spatial scale & 0.3 arcsec pixel$^{-1}$ & 0.3 arcsec pixel$^{-1}$ \\
Slit width & 1.04 arcsec & 1.04 arcsec \\
\hline
\end{tabular}
\end{table}

\begin{table*}
\begin{minipage}{13cm}
\caption{\label{obssample}Observations and galaxy sample}

\begin{tabular}{lccccccc}

\hline
Galaxy & Alternate & R.A. & Decl. & P.A. & Night & $\lambda_{c}$ & Exposure\\
& Designation & (1950) & (1950) &($^{\circ}$) && (\AA) & (s) \\
\hline

NGC 4235 & IC 3098  & 12 14 36 & + 07 28 09 & 70 & 2 & 5100 & 3600 \\
&&&&&& 8700 & 3600 \\  

NGC 5940 & UGC 9876 & 15 28 51 & + 07 37 38 & 70 & 2 & 5100 & 3600 \\
&&&&&& 8700 & 3600 \\

NGC 6104 & UGC 10309 & 16 14 40 & + 35 49 50 & 83 & 1 & 5100 & 3600 \\ 
&&&&&& 8700 & 3600 \\

Mk 270 & NGC 5283 & 13 39 41 & + 67 55 28 & 70 & 2 & 5100 & 3600 \\
&&&&&& 8700 & 3600 \\

Mk 423 & MCG 6-25-72 & 11 24 07 & + 35 31 34 & 170 & 1 & 5100 & 5400 \\
&&&&&& 8700 & 5400 \\

Mk 759 & NGC 4152 & 12 08 05 & + 16 18 41 & 35 & 2 & 5100 & 3600 \\
&&&&&& 8700 & 3600 \\

Mk 766 & NGC 4253 & 12 15 55 & + 30 05 27 & 108 & 1 & 5100 & 5400 \\
&&&&&& 8700 & 5400 \\

Mk 885 & --- & 16 29 43 & + 67 29 06 & 90 & 1 & 5100 & 3600 \\
&&&&&& 8700 & 3600 \\

Mk 1098 & --- & 15 27 37 & + 30 39 24 & 70 & 2 & 5100 & 3600 \\
&&&&&& 8700 & 3600 \\

3C305 & IC 1065 & 14 48 18 & + 63 28 36 & 57 & 1,2 & 5100 & 3600 \\
&&&&&& 8700 & 3600 \\

\hline

\end{tabular}
\end{minipage}
\end{table*} 


\section{Observations and data reductions}

High resolution long-slit spectra from the galaxies of our sample were 
obtained in 1991 May during two observing nights
with the WHT telescope at the Roque de los Muchachos Observatory, in
the Spanish island of La Palma. Details concerning
the set-up for the observations can be found in Table~\ref{obs}.
We used the two arms of the ISIS spectrograph
to take, simultaneously two spectra of each galaxy in different spectral 
ranges. We observed 11 galaxies. One of them (the radiogalaxy Hydra A)
is not included in this work because of the poor quality  of its spectra. 
Two different exposures of 30 min.(three in some objects, see 
Table~\ref{obssample}) were taken on each galaxy. This allowed us to clean 
the spectra of cosmic ray events. Beside the galaxies of the sample, we 
also observed several stars for flux calibration, subtraction of water 
vapour bands, and velocity measurements.

The reduction of the data was done using standard tools in the {\sc iraf} 
package. The reduction process was performed fully in two dimensions and 
consisted of several steps: bias subtraction,
flat-fielding, cosmic-rays cleaning, wavelength and flux calibration and,
finally, sky subtraction. This last step was the most difficult one, at least 
for the near--IR spectra since, in the spectral range covered by these 
spectra, there are many atmospheric absorption and emission bands.
The emission features were eliminated by subtracting spectra of the sky from 
those of the galaxies. The elimination of the OH and water absorption bands 
was achieved by dividing by the spectrum of a standard star. This procedure 
is not perfect. It has two main problems. The first one is that atmospheric 
conditions change during the night.
The second one is that the absorption lines of the star cannot be eliminated 
when they coincide with the atmospheric absorption lines. 
For this reason, when dividing the galaxy spectra by 
the stellar one, there often appear weak pseudo-emission lines that have 
their origin on
photospheric absorption lines in the stellar spectrum. In Fig.~\ref{comp}
we can see two examples of the results of the correction  procedure. In 
Mk~1098, observed during the first night, we can see that the original 
strong absorption bands appear, after division, like weak ``emission bands''. 
This can be due to changes in the atmospheric conditions during the night.
In Mk 423, observed during the second night, 
weak narrow ``emission lines'' can be seen due to absorption lines 
in the spectrum of the star.

One dimensional spectra were extracted at the end of the reduction process by 
adding the ten central pixels. This corresponds to 3~arcsec. The average 
seeing was about 1~arcsec.

The spectra were not corrected for galactic reddening.


\begin{figure*}
\centerline{
\psfig{figure=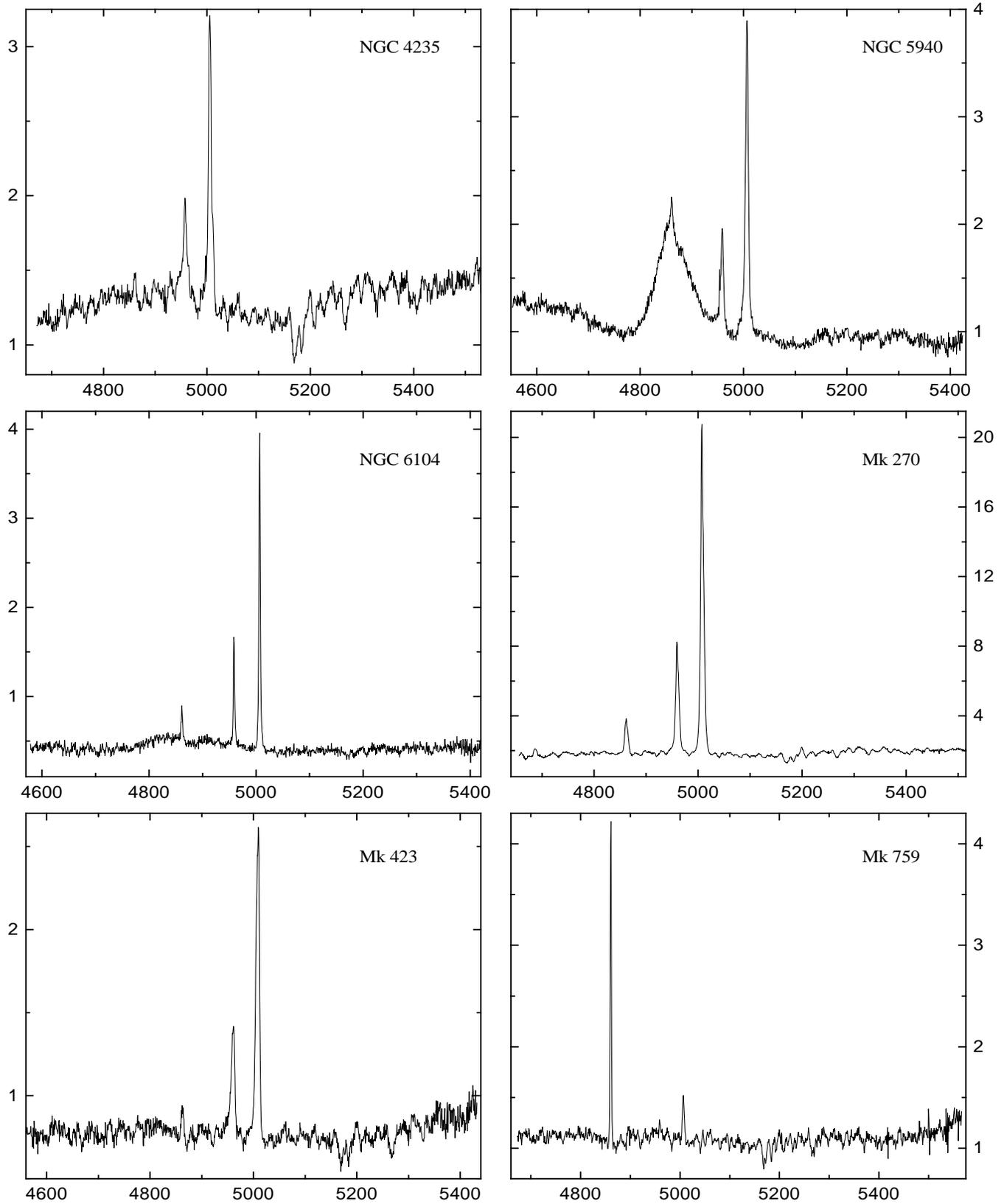,width=18cm,height=22cm,clip=}}
\caption{Blue spectra of the galaxies of the sample. The flux is
in units of 10$^{-15}$erg cm $^{-2}$ s$^{-1}$ and the wavelengths are in \AA\ .
Note that the scale in each spectrum is different.}
\label{littlespectra}
\end{figure*}

\addtocounter{figure}{-1}
\begin{figure*}
\centerline{
\psfig{figure=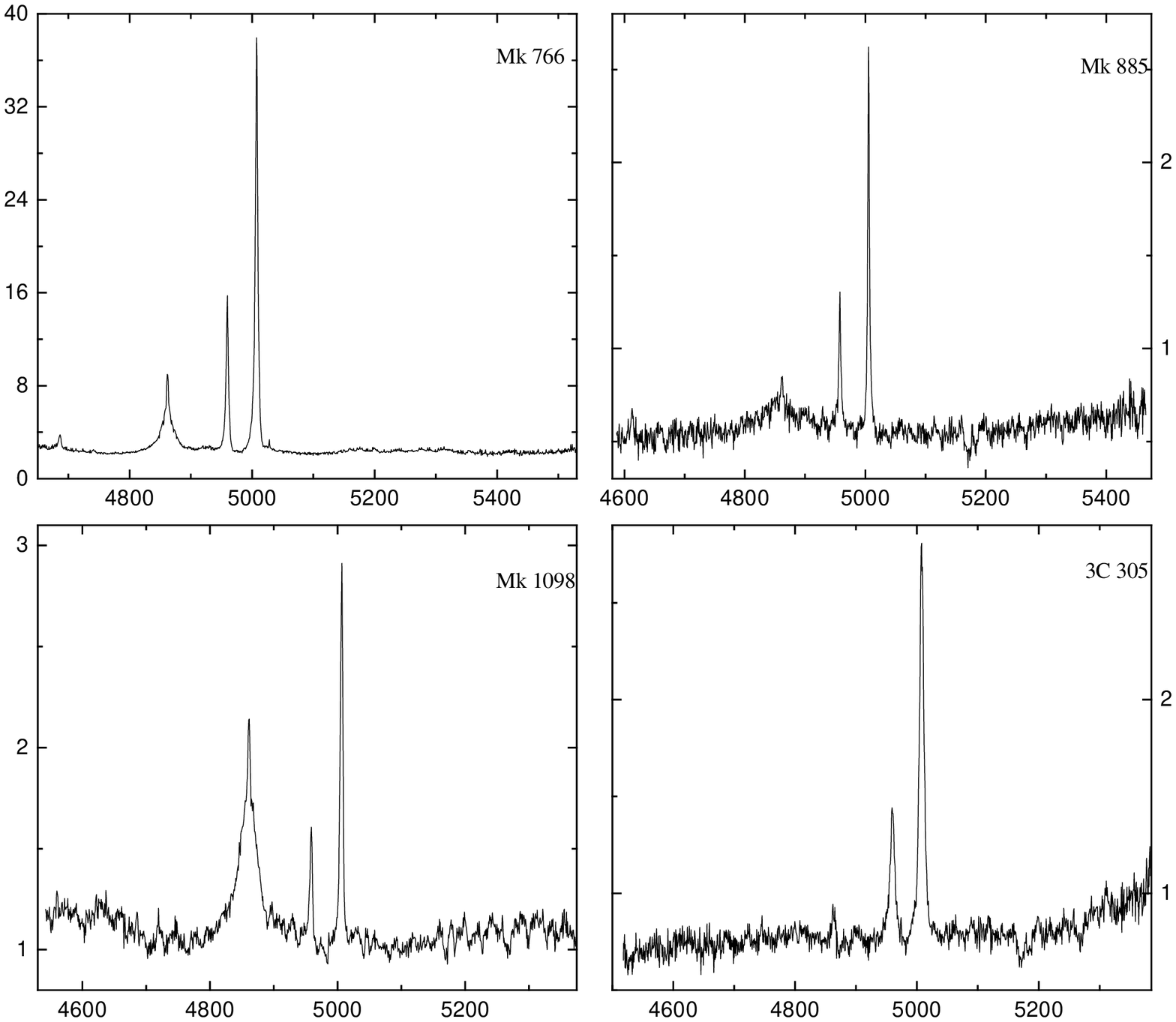,width=18cm,height=22cm,clip=}}
\caption{cont.}
\label{littlespectrab}
\end{figure*}

\begin{figure*}
\centerline{
\psfig{figure=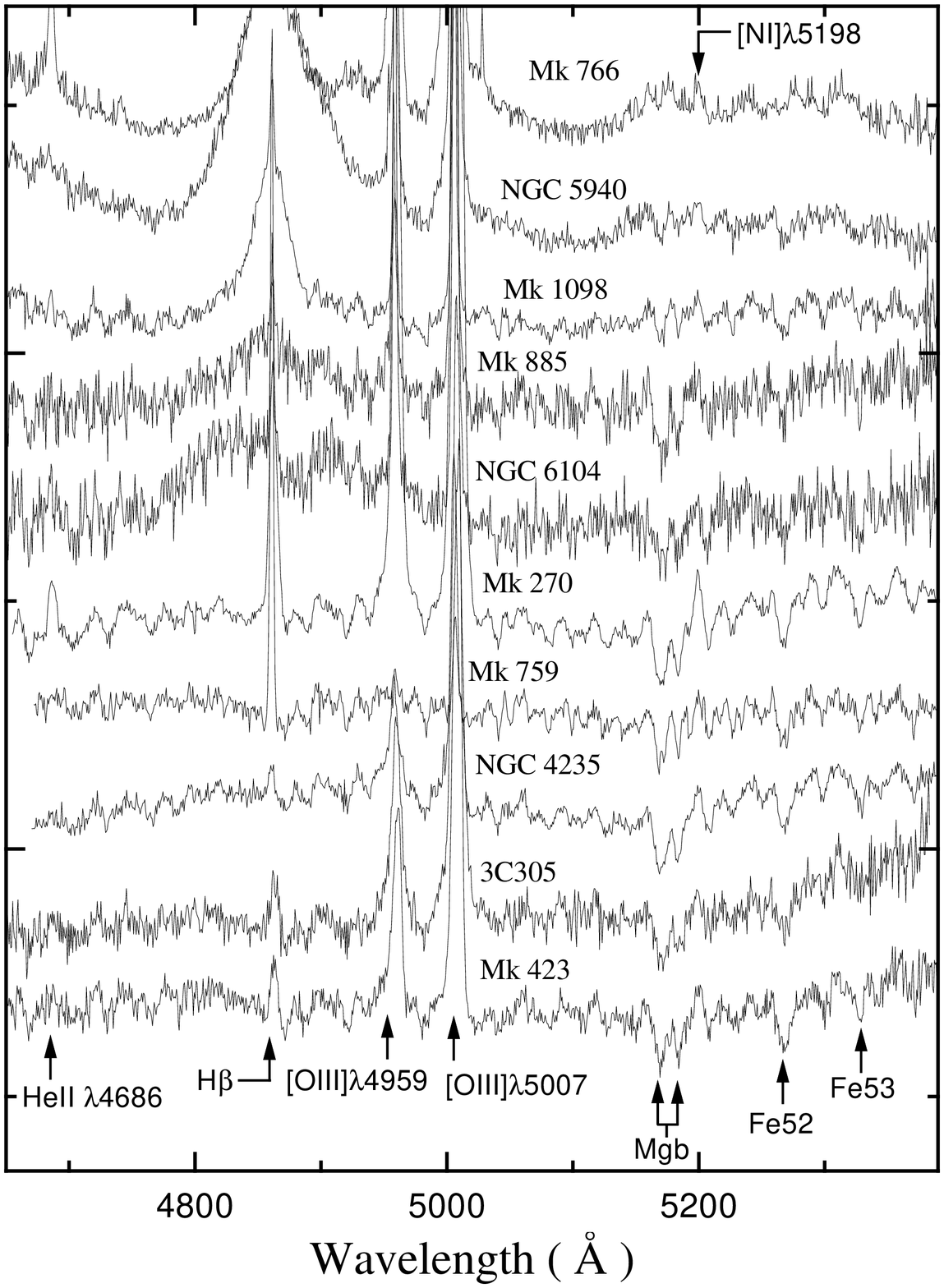,width=17cm,clip=}}
\caption{Enlarged blue spectra, showing the stellar absorption features of 
the galaxies of the sample}
\label{bluespectra}
\end{figure*}

\begin{figure*}
\centerline{
\psfig{figure=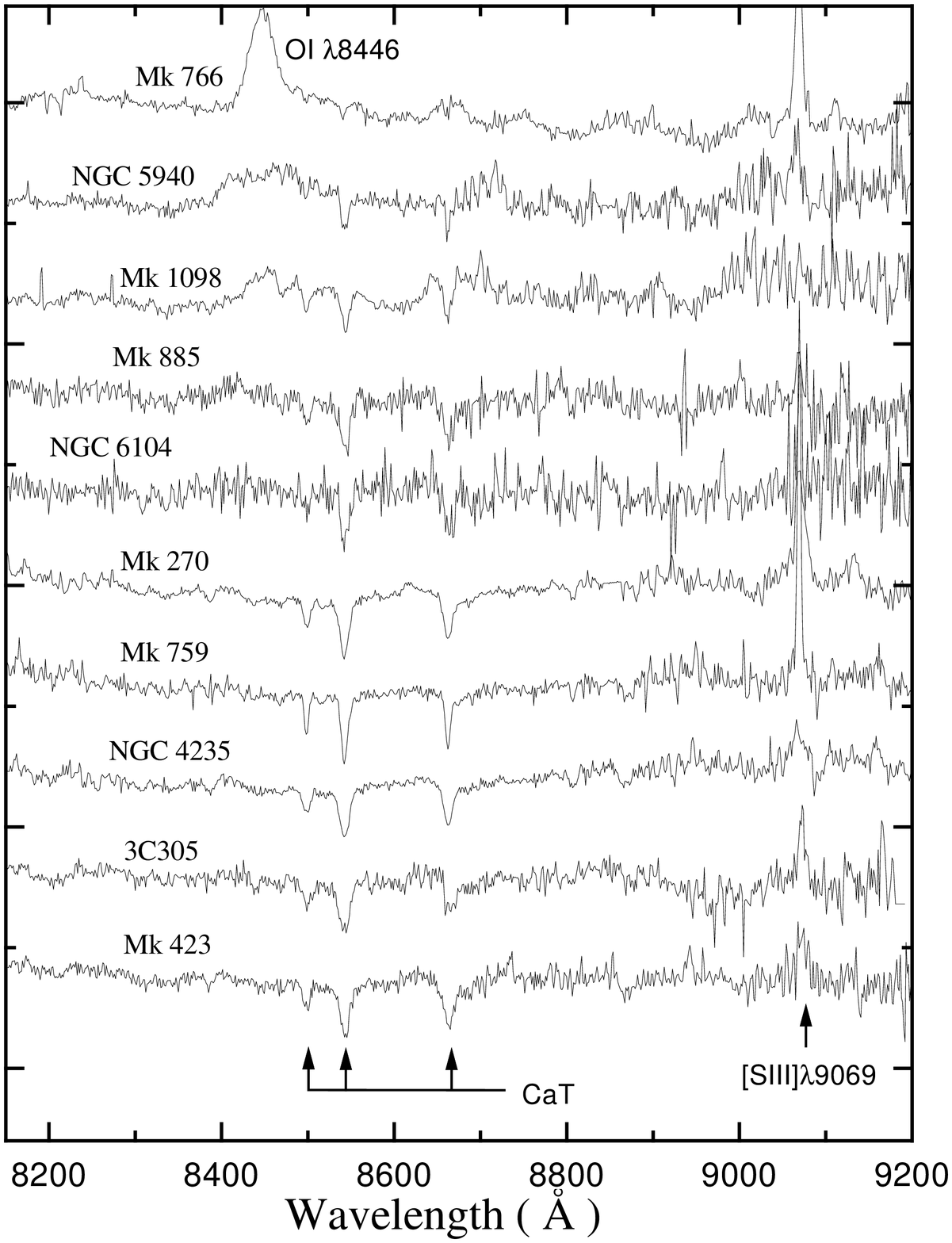,width=17cm,clip=}}
\caption{Enlarged near--IR spectra, showing the CaT. This feature is seen in 
absorption in all
galaxies, even in Mk~766, where CaT is also in emission.}
\label{redspectra}
\end{figure*}

\begin{table*}
\begin{minipage}{14cm}
\caption{\label{physsample}Physical properties of the sample. 
Col.~(2) gives the nuclear  type of the galaxy 
taken from the literature. 
Col~(3) tells us if the nucleus 
is spatially resolved or not in HST images from Malkan et~al. (1998).
The data of Cols.~(4)--(7) are from de Vaucouleurs et~al. (1991) (RC3). 
The values in Col.~(8) have been calculated with data in Cols.~(6) and (7).
We have derived values in Col.~(9) from the empirical formulae of Simien \& 
de Vaucouleurs (1986) and the data in Col.~(8).}

\begin{tabular}{lcccccccc}

\hline
Galaxy & Type & Resolved & Morph. &
$T$ & $cz$ & $B^o_T$ & $M_{B^o_T}$ & $M_{bulge}$ \\
\hspace{2mm} (1) & (2) & (3) & (4) & (5) & (6) & (7) & (8) & (9)\\
\hline

NGC 4235 
& Sy 1 & yes & SA(s)a & 1.0 $\pm$ 0.3 & 2343 $\pm$ 33 & 11.93 & -20.5 & -19.5 
$\pm$ 0.6 \\

NGC 5940  
& Sy 1 & no & SBab & 2.0 $\pm$ 0.9 & 10172 $\pm$ 21 & 14.13 & -21.5 & -20.3 
$\pm$ 2.8 \\

NGC 6104   
& Sy 1.5 & yes & S? & --- & 8382 $\pm$ 50 & 14.04 & -21.2 & --- \\

Mk 270  
& Sy 2 & yes & S0? & -2.0 $\pm$ 1.7 & 3165 $\pm$ 86 & 14.12 & -19.0 & -18.5 
$\pm$ 1.5 \\

Mk 423  
& Sy 1.9 & yes & S0? & --- & 9652 $\pm$ 11 & 14.63  & -20.9 & --- \\

Mk 759  
& \hbox{H\,{\sc ii}} & yes & SAB(rs)c & 5.0 $\pm$ 0.3 & 2066 $\pm$ 46 & 12.38 & -19.8 & -17.3
 $\pm$ 1.8 \\

Mk 766  
& Sy 1.5 & no & SB(s)a & 1.0 $\pm$ 0.6 & 3819 $\pm$ 25 & 13.74 & -19.8 & -18.8
 $\pm$ 1.3 \\

Mk 885  
&  Sy 1.5 & yes & S? & --- & 7593 $\pm$ 30 & 14.86 & -20.2 & --- \\

Mk 1098  
& Sy 1.5 & --- & S? & --- & 10710 $\pm$ 49 & --- & --- & --- \\

3C305  
&  NLRG & --- & SB0 & -2.0 $\pm$ 0.8 & 12423 $\pm$ 53 & 14.43 & -21.7 & -21.1 
$\pm$ 0.7\\
\hline

\end{tabular}
\end{minipage}
\end{table*} 


\section{Notes on individual objects}

The main physical properties of the 10 galaxies of the sample are given in 
Table~\ref{physsample}. The redshift of the sample covers a range 
0.007 -- 0.041 and therefore a factor of 4 in size. Their nuclear blue spectra 
can be seen in 
Figs.~\ref{littlespectra} and~\ref{bluespectra}. In this last figure,
the spectra are enlarged  in order to show the stellar features in the 
nuclei of the galaxies.
The near--IR spectra are shown in Fig.~\ref{redspectra}. The CaT is seen in 
absorption in all of the nuclei, even in that of Mk 766, where it is detected also 
in emission.

  \subsection{NGC 4235} 

NGC 4235 is classified as a Seyfert 1 galaxy in the literature 
(Abell, Eastmond \& Jenner 1978; Morris \& Ward 1988).
This is a nearly edge-on spiral galaxy and its nuclear emission is
strongly reddened by the dust in the spiral arms of the galaxy \cite{AEJ78}.
It has a spatially resolved nucleus, bisected by a dust lane \cite{MGT98}
and extended radio-emission spread on both sides of the galaxy \cite{Col96a}
with a compact radio core
(Ulvestad \& Wilson 1984b; Kukula et~al. 1995). 
The galaxy presents optical emission lines extended along the major axis
(Pogge 1989; Colbert et~al. 1996b).   

  \subsection{NGC 5940}
NGC 5940 is a barred spiral galaxy with an unresolved point source in the 
nucleus \cite{MGT98}.
It is a prototypical Seyfert 1 galaxy with broad permitted emission lines 
like H$\beta$ and \hbox{O\,{\sc i}} $\lambda$8446
and it is also a strong \hbox{Fe\,{\sc ii}} emitter. The multiplets of 
\hbox{Fe\,{\sc ii}} form two big bumps in the spectrum of NGC 5940, 
one around 4600\AA\ in the blue edge of 
the spectrum (see Fig.~\ref{bluespectra}), and the other one
between 5150\AA\ and 5250\AA. The first of these bumps is due to multiplets 
(37) and (38), while
the second one is due to multiplets (42), (48) and (49) \cite{Ost77}. 
Besides strong emission lines, many stellar absorption features can be seen 
in its spectra: the Mgb, Fe$_{52}$ and  Fe$_{53}$ features are
clearly visible. The near--IR spectrum of NGC 5940 is affected by 
problems with sky subtraction, and
the third line of the CaT has been 
nearly lost. A comparison with older spectra \cite{MW88}
indicates that the galaxy has not suffered violent
changes in the last years.

  \subsection{NGC 6104}
NGC 6104 shows a spatially resolved nucleus, a nuclear bar and
a strongly disturbed spiral pattern that can appear like a ring in low 
resolution images \cite{MGT98}. 
There are not good quality images of the external region; the image in the
Digital Sky Survey presents a highly disturbed appearance.
 
Our spectrum shows that it could be classified as
a Seyfert 1.5 to 1.8 galaxy. Most striking about this spectrum is the shape
of the spectral lines. The broad component of H$\beta$ shows  an asymmetrical 
double-peaked  shape and the forbidden
emission lines are very narrow, and are only marginally resolved. 
  
  \subsection{Mk 270}
This is a nearly face-on S0/E galaxy. It shows filaments, 
wisps and dust lanes in its inner regions \cite{MGT98}. 
It has extended \hbox{[O\,{\sc iii}]} and H$\alpha$ emission elongated at
P.A. = 58$^{\circ}$ (Haniff, Wilson \& Ward 1988; 
Mulchaey, Wilson \& Tsvetanov 1996). It has also two weak radio-components on both sides of the 
nucleus elongated in the same direction as the extended optical emission 
(Ulvestad \& Wilson 1984a; Kukula et~al. 1995).

  \subsection{Mk 423}
Mk 423 is classified in the literature as a type 1.9 Seyfert galaxy 
\cite{Ost81}. Our spectra confirm that H$\beta$ has no broad component. 

This galaxy is interacting with a nearby companion \cite{MGT98}. In the literature it
is often included among Markarian galaxies with multiple nuclei (e.g. Nordgren et~al. 1995).
Mk 423 and its companion have both extended emission line regions seen in 
\hbox{[O\,{\sc iii}]} and H$\alpha$ \cite{RCP85}. Our blue spectrum of the
companion suggests an  \hbox{H\,{\sc ii}} galaxy classification.
The radio core of Mk 423 is partially resolved \cite{Ulv86}. 

One  interesting
feature in the spectrum of this galaxy is the broad absorption wings of 
H$\beta$, suggesting the presence of young main sequence stars.
Rudy, Cohen \& Puetter  
\shortcite{RCP85} found a moderately strong UV continuum in 
the 1200--2000~\AA\ spectrum but no signs of broad emission lines.
This UV continuum emission could be produced by the same young stars that
produce the broad absorption wings of H$\beta$.

  \subsection{Mk 759}
Mk 759 is the only starburst galaxy in our sample, included for comparison 
purposes. 
H$\beta$ has broad absorption wings, indicating 
the presence of young stars.

  \subsection{Mk 766}
Mk 766 is a barred spiral galaxy. HST images \cite{MGT98} show 
filaments, 
wisps and irregular dust lanes around an unresolved nucleus. 
It has a partially resolved radio-core with an extension to the 
north.
The total size of the source is $\sim$ 200 pc $h^{-1}$ 
\cite{UW84a}. The polarization angle of 
the light emitted by Mk 766 is approximately perpendicular to
the radio axis \cite{Good89}.  The optical emission is extended 
(Gonz\'alez--Delgado 
\& P\'erez 1996; Mulchaey et~al. 1996) through a region whose total size is
greater than that of the radio-source. In X-rays, it is variable 
on a few hours time-scale, and presents a strong soft X-ray excess (Molendi, 
Maccacaro \& Schaeidt 1993). Molendi \& Maccacaro \shortcite{MM94} asserted 
that the soft 
X-rays excess and the hard X-ray emission are produced by two 
distinct mechanisms. Nandra et~al. \shortcite{Nan97} reported
the detection of a Fe K${\alpha}$ line in the X-ray spectrum.

It is classified as a narrow line Seyfert 1 galaxy \cite{OP85}.
 Much work has been done on its emission spectrum 
(e.g.~Veilleux 1991). The most thorough one is by Gonz\'alez--Delgado 
\& P\'erez \shortcite{GP96}. 
Mk 766 is the only galaxy in the sample that shows CaT in emission 
\cite{Per88}. 
Interestingly, in our spectrum, as in the one by Gonz\'alez Delgado \& P\'erez 
\shortcite{GP96},
the CaT is also seen in absorption (see Fig~\ref{redspectra}).

  \subsection{Mk 885}
Mk 885 is a barred spiral galaxy \cite{XR91}, classified as a 1.5 Seyfert 
galaxy. Our spectra of this galaxy are not very good, but we can confirm the
results from Osterbrock \& Dahari \shortcite{OD83} that the nuclear spectrum has
stellar absorption features. Its nucleus is spatially resolved \cite{MGT98} at
HST resolution.

  \subsection{Mk 1098}
There are no good images from this galaxy available and its morphological 
parameters are not well known. The nuclear spectrum of Mk 1098 looks like that of NGC 5940, 
but with permitted emission lines weaker and narrower than those of the latter. It
is a moderately strong \hbox{Fe\,{\sc ii}} emitter.
The most striking feature in the spectra of this
galaxy is the strange shape of the feature that forms the \hbox{O\,{\sc i}} 
$\lambda$8446 
emission line and the CaT. While part of this could be due to a poor sky subtraction (see 
Fig.~\ref{comp}),
the CaT might also be in emission in the nucleus.

  \subsection{3C 305}
This galaxy is the only radio galaxy in our sample.
Its blue spectrum is very similar to the
one of Mk 423. The results will be presented elsewhere.



\section{Measurements}

  \subsection{Line Strengths}

   \subsubsection{Definitions}
Our central interest is to measure the strength of optical and NIR stellar
absorption features. For that we follow TDT90 approach and measure 
`pseudo-equivalent widths'.

The definition of the pseudo-equivalent width of a spectral index is:
\[
W_{c} \equiv \Delta \lambda_{c} \left(1- \frac{F^{L}_{c}}{F^{C}_{c}}\right) 
\]
where $\Delta \lambda_{c}$ is the width of the central window and $F^{L}_{c}$ 
is the mean flux in that band-pass. $F^{C}_{c}$ is the value of the pseudo-continuum 
in the central wavelength $\lambda_{c}$.
The pseudo-continuum  is defined by
interpolation of the flux in two continuum band-passes:
\[
F^{C}_{c} \equiv \frac{F^{C}_{b} - F^{C}_{a}}{\lambda_b - \lambda_a}\lambda_c +
\frac{F^{C}_{a}\lambda_{b} - F^{C}_{b}\lambda_a}{\lambda_{b} - \lambda_{a}}
\]
where $F^{C}_{b}$ and $F^{C}_{a}$ are the mean values of the flux in the 
continuum windows
and $\lambda_a$ and $\lambda_b$ are their central wavelengths.
Hereafter, pseudo-equivalent widths will be called equivalent widths (EWs) 
for simplicity.

Table~\ref{bandpass} gives the 
definition of the relevant continuum and central band-passes. 
Those corresponding to the blue indices are taken from the 
Lick system (e.g. Gorgas et~al. 1993),
whereas those of the CaT are taken from TDT90.

The errors of the EWs were calculated using a simplified version of the 
photon statistic method.
The measured r.m.s. of the spectra in the continuum band-passes, were taken
as the errors of the
mean values of the continuum in these bands. The errors of the EWs were then 
calculated by propagating 
errors from the above expressions \cite{Pal97}. These errors are thus only lower limits.

\begin{table*}
\begin{minipage}{12.5cm}
\caption{\label{bandpass}Definitions of the atomic indices. 
The definition of the Mgb index is the one of the Lick system.
The Mgb opt. 1 and Mgb opt. 2 indices have optional red windows defined by
Palacios et~al. 1997.
The indices CaT 1,2 and 3 measure the
EW of each of the three lines of the CaT as defined by TDT90.}
\begin{tabular}{lccc}

\hline
Index & Continuum blue band-pass & Central band-pass & Continuum red 
band-pass \\
      &  (\AA)    &  (\AA)       &   (\AA)  \\

\hline
Fe$_{52}$ & 5235.50 -- 5249.25 & 5248.00 -- 5286.75
& 5288.00 -- 5319.25 \\

Fe$_{53}$ & 5307.25 -- 5317.25 & 5314.75 -- 5350.50
& 5356.00 -- 5364.75 \\ 

Mgb & 5144.50 -- 5162.00 & 5162.00 -- 5193.25
& 5193.25 -- 5207.00 \\

Mgb opt. 1 & 5144.50 -- 5162.00 & 5162.00 -- 5193.25
& 5207.00 -- 5221.00 \\

Mgb opt. 2 & 5144.50 -- 5162.00 & 5162.00 -- 5193.25
& 5220.00 -- 5234.00 \\

\\
CaT 1 & 8447.50 -- 8462.50 & 8483.00 -- 8513.00
& 8842.50 -- 8857.50 \\

CaT 2 & 8447.50 -- 8462.50 & 8527.00 -- 8557.00
& 8842.50 -- 8857.50 \\

CaT 3 & 8447.50 -- 8462.50 & 8647.00 -- 8677.00
& 8842.50 -- 8857.50 \\
\hline

\end{tabular}
\end{minipage}
\end{table*}

\subsubsection{Broadening correction}
The velocity dispersion of the stars broaden the absorption lines in the
spectra of galaxies; therefore the central band-passes of the indices may 
not contain the complete feature. A 
correlation exists between the velocity dispersion of a given galaxy and 
the EWs of the absorption lines in the integrated spectrum and a 
correction for this effect is needed. For more details see TDT90.

   \subsubsection{Emission line effects}
\label{problems}

The Mg$_2$ index is more affected than the Mgb by the presence of emisssion lines 
in AGNs, so we have measured Mgb.

The \hbox{[N\,{\sc i}]} $\lambda$5198 doublet inside the red continuum 
band-pass of the Mgb index still remains a problem. 
One method to solve it is to
eliminate the line of \hbox{[N\,{\sc i}]} \cite{GE96}; another one is 
to use optional band-passes. 
The correction factors for the optional definitions of the Mgb index are 
very close to unity. 

To measure the EW of the Mgb two optional band-passes, defined in Palacios et~al. \shortcite{Pal97} 
were used. Whithout applying any correction, the two measurements were averaged to find the EW of 
the Mgb index.
To test the validity of the method, the emission lines
in the spectra of two of the galaxies: NGC 4235 and Mk 270 were eliminated. Of all of the 
galaxies in the sample, these are the ones that have the oldest stellar 
population. Their spectrum looks like that of a red giant star. Three 
spectra  of K type stars, taken the same night as the spectra
of the galaxies, were combined to form a template calibration star. Using 
an iterative procedure \cite{CGA98} 
the ranges affected by emission lines in the spectrum of the galaxies were fitted and replaced.
The measurements obtained with this method in Mk 270 and NGC 4235 are 
the same as those obtained by using our optional windows.

In the near--IR spectra, the problems are due to the presence of the broad emission 
line \hbox{O\,{\sc i}} $\lambda$8446 in the blue continuum band-pass of the CaT 
indices, and to residuals of the sky subtraction in the red band-passes. 

Fortunately, the
broad emission \hbox{O\,{\sc i}} $\lambda$8446 was a problem only in two 
galaxies:
NGC 5940 and Mk 1098. For these, we fitted two gaussians to the broad component of 
H$\beta$ in both galaxies, and eliminated the \hbox{O\,{\sc i}} 
line by taking these gaussians as an initial fit. The \hbox{O\,{\sc i}} 
emission line in NGC 5940, could be properly subtracted with only small
modifications to the original gaussians. Mk 1098 \hbox{O\,{\sc i}} line
could only be fitted by one of the gaussians. 

The red band-passes of the CaT indices were manually cleaned in all of the 
spectra to eliminate the residuals of the sky subtraction.

Mk 766 is a special case. It shows many broad emission lines in the near--IR
(\hbox{O\,{\sc i}} $\lambda$8446 and some lines of the Paschen series; even
the CaT is in emission). The stellar features in the blue are also completely 
diluted. No stellar feature could therefore be measured in this galaxy.

   \subsubsection{Results}
The EWs of the different lines for the galaxies in the sample are given in 
Table~\ref{EWresults}, both corrected and non-corrected for broadening.
This correction is smaller than the error in all galaxies. 
In the analysis of the next section, the corrected values are used.
For Mgb we use the results obtained 
with optional band-passes. For consistency with previous work (TDT90), 
hereafter we call CaT index the sum of the EWs of the second and third 
lines of the CaT: CaT (2+3). The EW of the third line of the CaT in NGC 5940 
is only a lower limit due to the effect on  this line of the atmospheric 
absorption bands (see Fig.~\ref{redspectra}).

  \subsection{Velocity Dispersions}

   \subsubsection{Stellar velocity dispersions}
\begin{figure}
\centerline{
\psfig{figure=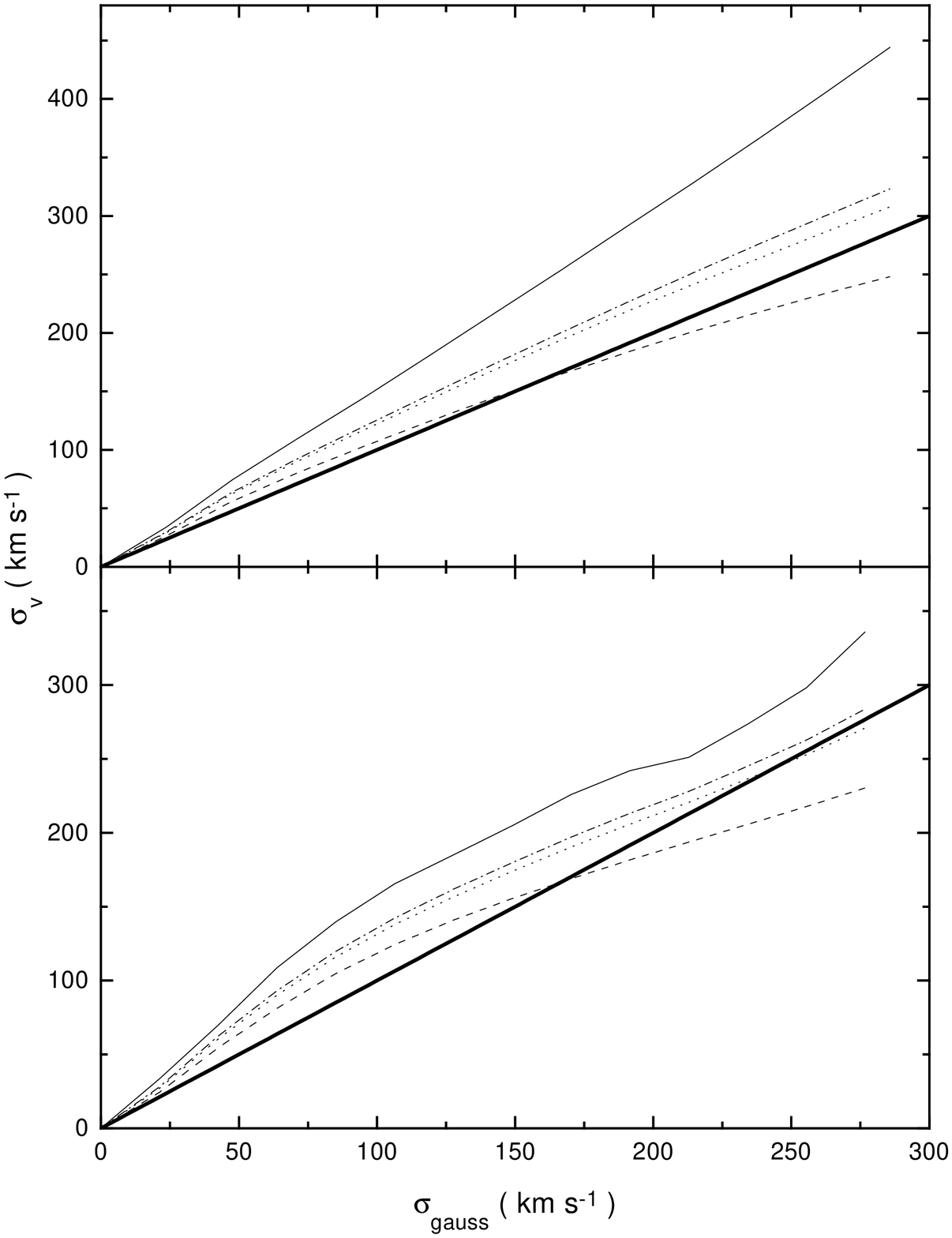,width=8cm,clip=}}
\caption{Velocity dispersion ($\sigma_{v}$) of a broadened stellar spectrum,
obtained by cross-correlation
with a stellar template vs. the widths of the gaussians used to convolve it ($\sigma_{gauss}$).
Values obtained from filtered functions 
(dotted and dashed-lines) are closer to the real widths
(thick straight line) than those from non filtered ones (thin line)
both for the near--IR (top) and blue (bottom) spectra.}
\label{diag}
\end{figure}

To measure stellar velocity dispersions, we used the
cross-correlation method described by Tonry \& Davis \shortcite{TD79}.  
This method gives small systematic errors when directly applied,
as can be seen in Fig.~\ref{diag}. In this figure we represent the velocity
dispersion computed for a broadened stellar spectrum applying the cross--correlation method,
vs. the width of the gaussian used to broaden it.
It is clear form the figure that, no matter how the correlation function is filtered,
the method always gives small errors.
To correct this effect, 6th order empirical correction curves 
were calculated (Nelson \& Whittle 1995; Palacios et~al. 1997).

Several correlation functions were calculated for every galaxy. 
We filtered each function in different ways, measured the width of the
respective central peaks and applied  the 
corresponding correction curve to each  measurement. 
Finally, the velocity dispersion of the galaxy was obtained by averaging all 
the measurements. 

Two independent measurements of the velocity dispersion of each galaxy were
obtained, using the blue and the near--IR spectra respectively.


\begin{table*}
\begin{minipage}{\textwidth}
\caption{\label{EWresults}EWs of the features in the galaxies of the
sample.}
\begin{tabular}{lccccccccc}
\hline
                        & NGC 4235 & NGC 5940 & NGC 6104 & Mk 270   & Mk 423  
 & Mk 759   & Mk 885   & Mk 1098  & 3C 305 \\
Index                   & EW (\AA) & EW (\AA) & EW (\AA) & EW (\AA) & EW (\AA)
 & EW (\AA) & EW (\AA) & EW (\AA) & EW (\AA)\\   
\hline

H$\beta_{broad}$     & $>$-35  & $\sim$ -85 & $\sim$ -40 & --- & --- & ---
& $\sim$ -35 & $\sim$ -45 & --- \\      

\hbox{[O\,{\sc iii}]} $\lambda$5007 & -13.6$\pm$0.9 & -26$\pm$1 & -35$\pm$2 & -81$\pm$4 &
        -23$\pm$1 & -2.0$\pm$0.1 & -17$\pm$1 & -10.6$\pm$0.6 & -27.7$\pm$0.6 \\

\\

Fe$_{52}$            & 2.7$\pm$0.3 & 0.7$\pm$0.3 & 1.6$\pm$0.6 & 2.5$\pm$0.3 
& 2.4$\pm$0.4 & 1.8$\pm$0.3 & 1.5$\pm$0.5 & 1.8$\pm$0.3 & 2.2$\pm$0.4\\
Fe$_{53}$            & 2.4$\pm$0.3 & 0.8$\pm$0.4 & 1.2$\pm$0.8 & 3.1$\pm$0.3 
& 2.3$\pm$0.5 & 1.2$\pm$0.4 & 1.9$\pm$0.8 & 1.1$\pm$0.3 & 1.8$\pm$0.7\\
Mgb                  & 4.6$\pm$0.3 & 0.9$\pm$0.3 & 2.7$\pm$0.6 & 5.3$\pm$0.5 
& 4.1$\pm$0.4 & 2.6$\pm$0.4 & 3.8$\pm$0.5 & 0.7$\pm$0.3 & 3.5$\pm$0.3\\
Mgb opt. 1           & 4.1$\pm$0.3 & 0.5$\pm$0.3 & 2.9$\pm$0.6 & 4.2$\pm$0.4 
& 3.3$\pm$0.4 & 2.1$\pm$0.3 & 3.0$\pm$0.5 & 0.3$\pm$0.3 & 3.1$\pm$0.3\\
Mgb opt. 2           & 4.2$\pm$0.2 & 0.3$\pm$0.3 & 3.0$\pm$0.5 & 4.2$\pm$0.3 
& 3.2$\pm$0.3 & 2.2$\pm$0.3 & 3.2$\pm$0.5 & 0.2$\pm$0.3 & 3.1$\pm$0.3\\

CaT (1)               & 1.3$\pm$0.1 & 1.3$\pm$0.2 & 1.3$\pm$0.3 & 1.3$\pm$0.1 
& 0.9$\pm$0.2 & 1.2$\pm$0.1 & 1.5$\pm$0.4 & 2.3$\pm$0.2 & 1.5$\pm$0.2\\
CaT (2)               & 3.3$\pm$0.1 & 1.7$\pm$0.2 & 2.9$\pm$0.3 & 3.7$\pm$0.1 
& 3.3$\pm$0.2 & 3.3$\pm$0.1 & 3.7$\pm$0.4 & 2.0$\pm$0.2 & 3.6$\pm$0.2\\
CaT (3)               & 2.9$\pm$0.2 & 1.3$\pm$0.2 & 2.6$\pm$0.3 & 2.9$\pm$0.1 
& 3.3$\pm$0.3 & 2.7$\pm$0.1 & 3.3$\pm$0.4 & 2.2$\pm$0.2 & 2.2$\pm$0.2\\
CaT (2+3)             & 6.3$\pm$0.3 & 3.0$\pm$0.4 & 5.5$\pm$0.6 & 6.6$\pm$0.2 
& 6.6$\pm$0.5 & 6.0$\pm$0.2 & 7.0$\pm$0.7 & 4.2$\pm$0.3 & 5.8$\pm$0.4\\

\\

Fe$_{52}$ ($\sigma=0$) & 2.9$\pm$0.4 & 0.8$\pm$0.3 & 1.7$\pm$0.8 & 2.7$\pm$0.4 
& 2.6$\pm$0.5 & 1.8$\pm$0.3 & 1.6$\pm$0.6 & 1.8$\pm$0.3 & 2.5$\pm$0.8\\
Fe$_{53}$ ($\sigma=0$) & 2.6$\pm$0.3 & 0.8$\pm$0.4 & 1.3$\pm$0.9 & 3.3$\pm$0.4 
& 2.5$\pm$0.5 & 1.2$\pm$0.5 & 2.0$\pm$0.8 & 1.1$\pm$0.3 & 2.0$\pm$0.8\\
Mgb ($\sigma=0$)     & 5.0$\pm$0.4 & 1.0$\pm$0.4 & 3.0$\pm$0.6 & 5.8$\pm$0.5 
& 4.5$\pm$0.5 & 2.7$\pm$0.4 & 4.1$\pm$0.5 & 0.7$\pm$0.4 & 4.0$\pm$0.5\\
Mgb opt. 1 ($\sigma=0$) & 4.4$\pm$0.4 & 0.5$\pm$0.3 & 3.1$\pm$0.6 & 4.4$\pm$0.4
& 3.6$\pm$0.4 & 2.1$\pm$0.3 & 3.2$\pm$0.6 & 0.3$\pm$0.3 & 3.4$\pm$0.4\\
Mgb opt. 2 ($\sigma=0$) & 4.5$\pm$0.3 & 0.4$\pm$0.3 & 3.2$\pm$0.6 & 4.4$\pm$0.4
& 3.4$\pm$0.4 & 2.2$\pm$0.3 & 3.4$\pm$0.5 & 0.2$\pm$0.3 & 3.4$\pm$0.4\\

CaT (1) ($\sigma=0$)  & 1.3$\pm$0.2 & 1.3$\pm$0.2 & 1.3$\pm$0.4 & 1.3$\pm$0.1 
& 0.9$\pm$0.2 & 1.2$\pm$0.1 & 1.5$\pm$0.4 & 2.2$\pm$0.2 & 1.5$\pm$0.2\\
CaT (2) ($\sigma=0$)  & 3.4$\pm$0.2 & 1.7$\pm$0.2 & 3.0$\pm$0.4 & 3.8$\pm$0.1 
& 3.5$\pm$0.3 & 2.8$\pm$0.1 & 3.8$\pm$0.4 & 2.0$\pm$0.2 & 3.8$\pm$0.3\\
CaT (3) ($\sigma=0$)  & 3.1$\pm$0.2 & 1.3$\pm$0.2 & 2.7$\pm$0.4 & 3.1$\pm$0.1 
& 3.5$\pm$0.4 & 2.8$\pm$0.1 & 3.5$\pm$0.4 & 2.3$\pm$0.2 & 2.5$\pm$0.3\\
CaT (2+3) ($\sigma=0$) & 6.5$\pm$0.3 & 3.0$\pm$0.4 & 5.7$\pm$0.7 & 6.8$\pm$0.2 
& 7.0$\pm$0.6 & 6.1$\pm$0.2 & 7.3$\pm$0.8 & 4.3$\pm$0.4 & 6.2$\pm$0.5\\

\hline

\end{tabular}
\end{minipage}
\end{table*}

\begin{table*}
\begin{minipage}{\textwidth}
\caption{\label{sigmagasblue}Blue emission line widths.}
\begin{tabular}{lcccccccc}
\hline
 & \multicolumn{3}{c}{------------ \hbox{[O\,{\sc iii}]} 
$\lambda$5007 ------------}
 &
\multicolumn{3}{c}{------------ \hbox{[O\,{\sc iii}]} 
$\lambda$4959 ------------}
& \multicolumn{2}{c}{--------- H$\beta$ ---------}\\

Galaxy & $\sigma_{0h}$ & $\sigma_{0.2h}$ & $\sigma_{0.5h}$ & $\sigma_{0h}$ 
& $\sigma_{0.2h}$ &$\sigma_{0.5h}$ & 
$\sigma_{narrow}$ & $\sigma_{broad}$ \\

       & (km s$^{-1}$) & (km s$^{-1}$) & (km s$^{-1}$) & (km s$^{-1}$) & 
(km s$^{-1}$) & (km s$^{-1}$) & (km s$^{-1}$) & (km s$^{-1}$)\\

\hline

NGC 4235 & \ctabla{196}{3} & \ctabla{185}{8} & \ctabla{151}{10} & 
\ctabla{226}{25} & \ctabla{146}{18} & \ctabla{131}{24} & ---  & ---    \\

NGC 5940 & \ctabla{169}{4} & \ctabla{167}{6} & \ctabla{151}{8} & 
\ctabla{213}{5} & \ctabla{220}{17} & \ctabla{169}{19} & \ctabla{260}{100} & 
\rtabla{1960}\\

NGC 6104 & \ctabla{64}{4} & \ctabla{60}{6} & \ctabla{57}{7} & \ctabla{61}{6} &
 \ctabla{55}{12} & \ctabla{49}{16} & \ctabla{91}{29} & \rtabla{3940}\\

Mk 270   & \ctabla{178}{2} & \ctabla{174}{3} & \ctabla{155}{3} & 
\ctabla{188}{6} & \ctabla{172}{4} & \ctabla{165}{5} & \ctabla{185}{3} & --- \\

Mk 423   & \ctabla{198}{3} & \ctabla{204}{8} & \ctabla{210}{13} & 
\ctabla{224}{13} &\ctabla{200}{21} & \ctabla{192}{33} & \ctabla{159}{7} & ---\\

Mk 759   & \ctabla{107}{6} & \ctabla{107}{23} & \ctabla{103}{31} & ---    & 
---         & ---         & \ctabla{49}{6 } & ---         \\

Mk 766   & \ctabla{142}{8} & \ctabla{132}{3} & \ctabla{106}{4} & 
\ctabla{126}{5} & \ctabla{122}{4} & \ctabla{90}{6} & \ctabla{121}{13} & 
\rtabla{785} \\

Mk 885   & \ctabla{97}{4} & \ctabla{83}{8} & \ctabla{66}{10} & \ctabla{119}{8}
 & \ctabla{93}{17} & \ctabla{68}{22} & \ctabla{217}{69} & 
\rtabla{2565} \\

Mk 1098  & \ctabla{121}{5} & \ctabla{123}{7} & \ctabla{115}{10} & 
\ctabla{128}{8} & \ctabla{138}{22} & \ctabla{114}{21} & \ctabla{164}{49} & 
\rtabla{1210}\\

3C305    & \ctabla{231}{4} & \ctabla{224}{8} & \ctabla{206}{13} & 
\ctabla{271}{9} & \ctabla{242}{22} & \ctabla{210}{33} & ---    & ---   \\

\hline

\end{tabular}
\end{minipage}
\end{table*}

    \subsubsection{Gas velocity dispersions}

The width of the emission lines H$\beta$, \hbox{[O\,{\sc iii}]} 
$\lambda \lambda$5007, 4959 and \hbox{[S\,{\sc iii}]} $\lambda$9069 were
measured to determine the velocity dispersion of the gas in the NLR. Three 
different methods were used:

\begin{enumerate}
\item 
Three times a suitable continuum was chosen by visual inspection and a
gaussian fitted to the whole line. The average of the measurements was 
obtained and the error calculated as the dispersion of these measurements, 
thus it is associated to the continuum determination. 
The width so obtained, we called $\sigma_{0h}$.
\item 
A second value ($\sigma_{0.2h}$)
 was obtained by fitting a gaussian, using
in the fitting only those points over $0.2h$ ($h$= total height of the
line above the continuum). 
\item 
The third one ($\sigma_{0.5h}$) was obtained by fitting the core of the line, but using 
only the points over $0.5h$.
\end{enumerate}
The line width was calculated as,

\[
\sigma_{g} = \sqrt{ \sigma^{2} - \sigma_{i}^{2} }
\]
where $\sigma$ is the dispersion of the gaussian fitted to each line and 
$\sigma_{i}$ is the instrumental width obtained from measurements of sky emission lines.

The pure emission spectrum was extracted for NGC 4235 and Mk 270 (see 
Sec.~\ref{problems}), and the emission line widths were measured on them.
In the Seyfert 1 and 1.5 of the sample, the widths of the broad ($\sigma_{b}$)
and narrow ($\sigma_{n}$) components of H$\beta$ were measured.

    \subsubsection{Results}
The measured emission line widths are shown in Tables~\ref{sigmagasblue} 
and~\ref{sigmagasred}. The widths of the line cores are used in the discussion
(i.e. $\sigma_{0.2h}$).

The stellar velocity dispersions (from the Mgb and CaT stellar features)
are given in Table~\ref{sigmastars}. Column 4 shows the average value of 
both measurements ($\sigma_*$). Although it doesn't have a clear physical 
meaning as the Mgb and CaT features can be produced by different stellar 
populations with different kinematical behaviour, the measurements 
obtained from the Mgb and from the CaT
are consistent in all galaxies, except in Mk 1098, 3C 305 and, marginally, 
in Mk 423. In these three galaxies
our measurements point to the presence of, at least, two distinct populations 
in their nuclei. 



\section{Discussion}

  \subsection{Dilution of the nuclear spectral indices}

In active galaxies, the EWs of the blue-optical stellar absorption features 
are smaller than those found  in non-active galaxies of the same morphological 
type. This effect has been called ``dilution'' in the literature, assuming 
that it is caused
by the BFC emitted by the AGN. Nevertheless, a nuclear featureless continuum 
is not necessary
to explain why the EWs of the blue indices are small in active galaxies.
According to stellar population models, the blue indices of a young or/and low metallicity
stellar population are also much smaller than those of the high metallicity nuclei of early 
type galaxies without active star formation. In this case, of course, the word ``dilution'' 
is not suitable, since the indices are not diluted by any BFC but they are intrinsically weak.

Nevertheless, we can define the effective dilution of an absorption feature in a galactic spectrum
as the ratio between the EW
of the feature in that galaxy and the average EW of the index in normal 
galaxies of the same morphological type:
\[
D_{Index}=\frac{EW_{Index}}{\langle EW_{Index} \rangle}
\]
This effective dilution is not real, in the sense that it has not
be caused by a power-law featureless continuum, but it is useful to quantify the 
different effects that reduce the EWs of the spectral indices of a galaxy.

In summary,
the dilution of the optical stellar features in Seyfert galaxies
is explained in different ways:

\begin{table}
\caption{\label{sigmagasred} Emission line widths in the near--IR spectra.}
\begin{tabular}{lccc}
\hline
 & \multicolumn{3}{c}{------------ \hbox{[S\,{\sc iii}]} 
$\lambda$9069 ------------ } \\

Galaxy & $\sigma_{0h}$ & $\sigma_{0.2h}$ & $\sigma_{0.5h}$ \\

       & (km s$^{-1}$) & (km s$^{-1}$) & (km s$^{-1}$) \\

\hline

NGC 4235 & \ctabla{232}{13} & --- & --- \\
 
NGC 5940 & --- & --- & --- \\

NGC 6104 & \ctabla{71}{6} & --- & --- \\

Mk 270   & \ctabla{202}{11} & \ctabla{201}{13} & \ctabla{178}{15} \\

Mk 423   & \ctabla{238}{24} & --- & --- \\

Mk 759   & \ctabla{36}{9} & \ctabla{42}{11} & \ctabla{38}{12} \\

Mk 766   & \ctabla{114}{3} & \ctabla{113}{8} & \ctabla{99}{10} \\

Mk 885   & \ctabla{125}{18} & \ctabla{98}{60} & --- \\

Mk 1098  & --- & --- & --- \\
 
3C 305   & \ctabla{201}{33} & --- & --- \\

\hline

\end{tabular}
\end{table}

\begin{table}
\caption{\label{sigmastars}Nuclear stellar velocity dispersions.}
\begin{tabular}{lccc}
\hline
Galaxy & $\sigma_{*}(Mgb)$ &$\sigma_{*}(CaT)$ & $\langle \sigma_{*} \rangle$ \\

       & (km s$^{-1}$) & (km s$^{-1}$) & (km s$^{-1}$) \\

\hline

NGC 4235 & \ctabla{160}{13} & \ctabla{153}{8} & \ctabla{155}{10} \\
 
NGC 5940 & \ctabla{112}{26} & \ctabla{93}{27} & \ctabla{103}{27} \\

NGC 6104 & \ctabla{144}{23} & \ctabla{135}{15} & \ctabla{138}{18} \\

Mk 270   & \ctabla{137}{11} & \ctabla{139}{8} & \ctabla{138}{10} \\

Mk 423   & \ctabla{167}{17} & \ctabla{151}{11} & \ctabla{156}{13} \\

Mk 759   & \ctabla{65}{10} & \ctabla{70}{8} & \ctabla{68}{9} \\

Mk 766   & \ctabla{106}{40} & --- & \ctabla{106}{40} \\

Mk 885   & \ctabla{145}{16} & \ctabla{144}{13} & \ctabla{144}{15} \\

Mk 1098  & \ctabla{77}{13} & \ctabla{122}{15} & \ctabla{96}{23} \\
 
3C 305   & \ctabla{212}{24} & \ctabla{170}{13} & \ctabla{180}{21} \\

\hline

\end{tabular}
\end{table}

\begin{enumerate}
\item In the {\it standard}  model, the continuum emitted by the nucleus 
of a Seyfert galaxy, is dominated by the featureless continuum
emission from the AGN that dilute the 
stellar spectral indices from the bulge stars.
\item 
If there are luminous starbursts 
in the central region of a Seyfert galaxy the EWs of the
blue absorption features like the Mgb or the Fe indices, 
will be weak like in the spectra of hot young stars.
 Therefore the observed optical continuum would be fairly featureless 
as it happens in starburst galaxies (TDT90; Cid Fernandes \& Terlevich 1995).
In the case of Seyfert 1 galaxies, the
broad emission lines of the BLR represent an extra source of dilution 
that must be taken into account.
\item 
In general, the lower the metallicity of a galaxy, the weaker its stellar 
features, the smaller the EWs. It is unlikely that the central regions of an 
early type galaxy, like the ones which commonly host Seyfert nuclei, are of 
low-metallicity so we are not going to consider this third possibility any 
further.
\end{enumerate}
There are several methods to discriminate which of these effects causes 
the dilution in the blue--optical spectrum of a certain galaxy. 
We have followed TDT90 and compared
the nuclear effective dilution of the Mgb index ($D_{Mgb}$) with the 
observed effective dilution of the CaT ($D_{CaT}$). If the light
emitted by the AGN is responsible for the dilution, $D_{blue}$ and $D_{CaT}$ 
would follow a simple relation, since the emission from the AGN follows a
power-law: $F_{\lambda} \propto \lambda^{-\alpha}$ (with  $0.8 \ga \alpha \ga 
1.7$). 
On the other hand, if the blue indices are weak because of the presence of young stars
in the nucleus, the CaT could be strong, since the CaT feature 
is strong in young red supergiants.

To measure the effective dilution of the stellar features in the spectra of 
Seyfert galaxies a reference value for the EWs of these features in normal 
non-active galaxies is needed. Unfortunately, such a reference value is not 
well established. Galaxian spectral indices depend on the age and metallicity 
of the stellar populations in that galaxy; therefore the EWs of the indices 
in normal galaxies show a wide range of values. To limit that range we must
choose template galaxies of the same type as the ones to study. Strictly 
speaking, templates should have the same red-shift and morphological properties
(Hubble type, bulge magnitude, etc.) as the observed galaxies 
but this is not always possible. Since Seyfert
galaxies are preferentially found among early type spirals \cite{HFS97},
we have searched the literature for samples of these galaxies with spectral 
indices measured. Reference values for the blue indices were obtained from
Idiart, de Freitas Pacheco \& Costa \shortcite{IFC96} and Fisher, Franx \& 
Illingworth \shortcite{FFI96}, chosen because their sample contain a 
substantial number of early type non-active spiral galaxies, and because the 
methods they used to measure the EWs are the same as ours. We have selected 
those galaxies with Hubble types between Sb ($T=3$) and E/S0 ($T=-3$). A 
total number of 62 galaxies were used. The reference values were calculated 
taking the median of the EWs of the lines in these galaxies and are 
shown in Table~\ref{EWref}. Using the larger (more than 500 galaxies)
--but also much more inhomogeneous-- sample of Bica et~al. \shortcite{Bic91}, 
we obtain a similar average value for Mgb (3.9 \AA).

\begin{table}
\caption{\label{EWref}Reference values for the atomic indices.}
\begin{tabular}{cccc}
\hline
Mgb     & Fe$_{52}$    & Fe$_{53}$    & CaT (1+2) \\
EW(\AA) & EW(\AA) & EW(\AA) & EW(\AA)   \\
\hline
3.8 $\pm$ 0.8 & 2.8 $\pm$ 0.6 & 2.6 $\pm$ 0.6 & 7.6 $\pm$ 0.8 \\
\hline
\end{tabular}
\end{table}

To find the average value of the EW of the CaT in normal spirals, we chose
from TDT90 those galaxies with $-3 \leq T \leq 
3$, and obtain, taking the median, a reference value of $7.6 \pm 0.8$. 
This value is more uncertain than the one for Mgb, because we only used 8 
galaxies to calculate it. Theoretical models (Garc\'{\i}a--Vargas, 
Moll\'a \& Bressan 1998) predict that the CaT of an old ($t \geq $ 2 Gyr)
single stellar population of solar metallicity should have an EW between 7 \AA\ 
and 9 \AA\, depending on the model chosen, which agrees with our reference value.

Although we believe that these average values are reasonably good, there is a 
large scatter in the EWs of the indices in normal early type spirals.
The EW of the Mgb index can be as low as  2 \AA, or as high as 5 \AA. Mgb EWs 
higher than 5 \AA\ are not common in spirals but they are in ellipticals. 
The range of values covered by the EW(CaT) is not so wide, usually 
between 6 and 8.5 \AA\ in spirals.

\begin{figure}
\centerline{
\psfig{figure=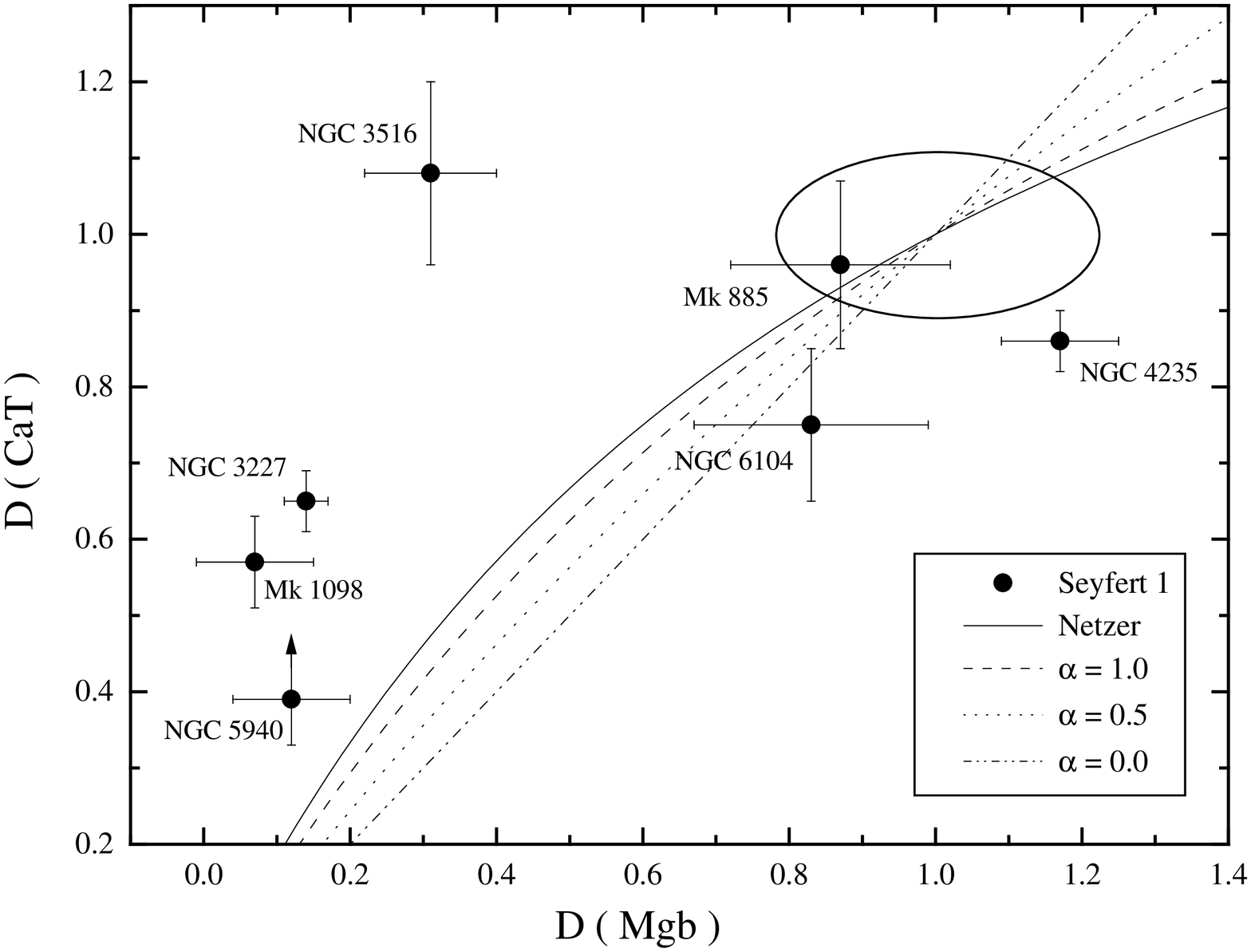,width=8.6cm,clip=}}
\caption{Dilution of the CaT  vs. dilution of the Mgb indices. The lines 
represent different power laws for the continuum $F_{\lambda} \propto 
\lambda^{-\alpha}$. 
The solid line represents a model (Netzer 1990) with two values of 
$\alpha$, one for the blue, another one for the near--IR. The ellipse shows 
the range of values covered by early type galaxies. The galaxies inside or 
near the ellipse are, therefore, not diluted with respect to normal spirals.}
\label{dilucion}
\end{figure}

$D_{Mgb}$ vs. $D_{CaT}$ are presented in Fig.~\ref{dilucion}.
Five Seyfert 1 galaxies of the present sample and two Seyfert 1 from TDT90 
sample  have been included in this figure. While TDT90 measured the CaT for 
these two galaxies, the Mgb value is from Dahari \& De Robertis 
\shortcite{DR89}. We can see the galaxies separating in two different groups:
in the first one (group I: NGC~4235, NGC~6104 and Mk~885) the EWs fall into 
the range of values covered by the indices of normal spiral bulges. 
These galaxies also share other properties:
\begin{enumerate}
\item The emission from the BLR is weak, with EW(H$\beta_{broad}$) $\ga -40$ \AA .
\item They are not  \hbox{Fe\,{\sc ii}} emitters.
\item Their nuclei are spatially resolved \cite{MGT98}.
\end{enumerate}
In summary the Seyfert activity in these galaxies appears to be weak.
Our spectra, because of their poor spatial resolution, include, not only the 
nuclei of the galaxies, but also, at least partially, their bulges. 
For these reasons we believe that, in these galaxies, the emission from the 
old stellar population in the bulge is masking the continuum 
emission from the active nucleus.

The galaxies of the second group (group II: NGC~3516, NGC~3227, NGC~5940 and Mk~1098)
are much more active: they have a prominent BLR with EW(H$\beta_{broad}$) $\la -45$ \AA , 
they are very strong  \hbox{Fe\,{\sc ii}} emitters and their nuclei are not spatially 
resolved. The spectral indices of
these galaxies, both the blue ones and the CaT, are diluted showing the contribution of an additional 
continuum over that corresponding to a ``normal''
galaxy nucleus. However, the 
relation between the dilutions of the blue indices
and the CaT does not conform to the predictions of the standard model, in which 
the continuum
emitted by the AGN follows a power-law. In these four galaxies the dilution 
of the blue indices is relatively greater
than that of the CaT. There are two possible explanations for this: the presence of a 
compact nuclear starburst in
the nuclei of these galaxies and/or the dilution of their blue 
indices by  \hbox{Fe\,{\sc ii}} multiplet emission.

NGC 3516 seems to be the most extreme case: only the blue indices are diluted. The CaT is 
instead slightly enhanced with respect to the values found in normal galaxies. 
Nevertheless, other observations \cite{Ser96} suggest that
the CaT may be somewhat diluted in the nucleus of NGC 3516, at least if we take 
as reference value the EW of the CaT in the bulge of this galaxy.

The redshift range covered by our sample (0.007 -- 0.041) implies different total areas for the fixed
10 pixel extraction in each galaxy, between 0.08 and 1.7 kpc$^2$. The results presented 
here are nevertheless barely affected by this, since the Seyfert 1 nuclear
luminosity dominates the spectra as evidenced by the fact that the BLR shows
up prominently even in the outermost pixels of the extraction windows. In any case, 
contamination by bulge stars would tend to decrease the dilution of the Mgb 
and  the CaT in the same proportion.

Circumnuclear star-forming regions cannot be ruled out in many Seyfert 1s since
it is not possible to resolve the inner 200--500 pc with ground based 
telescopes due to their large distance.

\begin{figure}
\centerline{
\psfig{figure=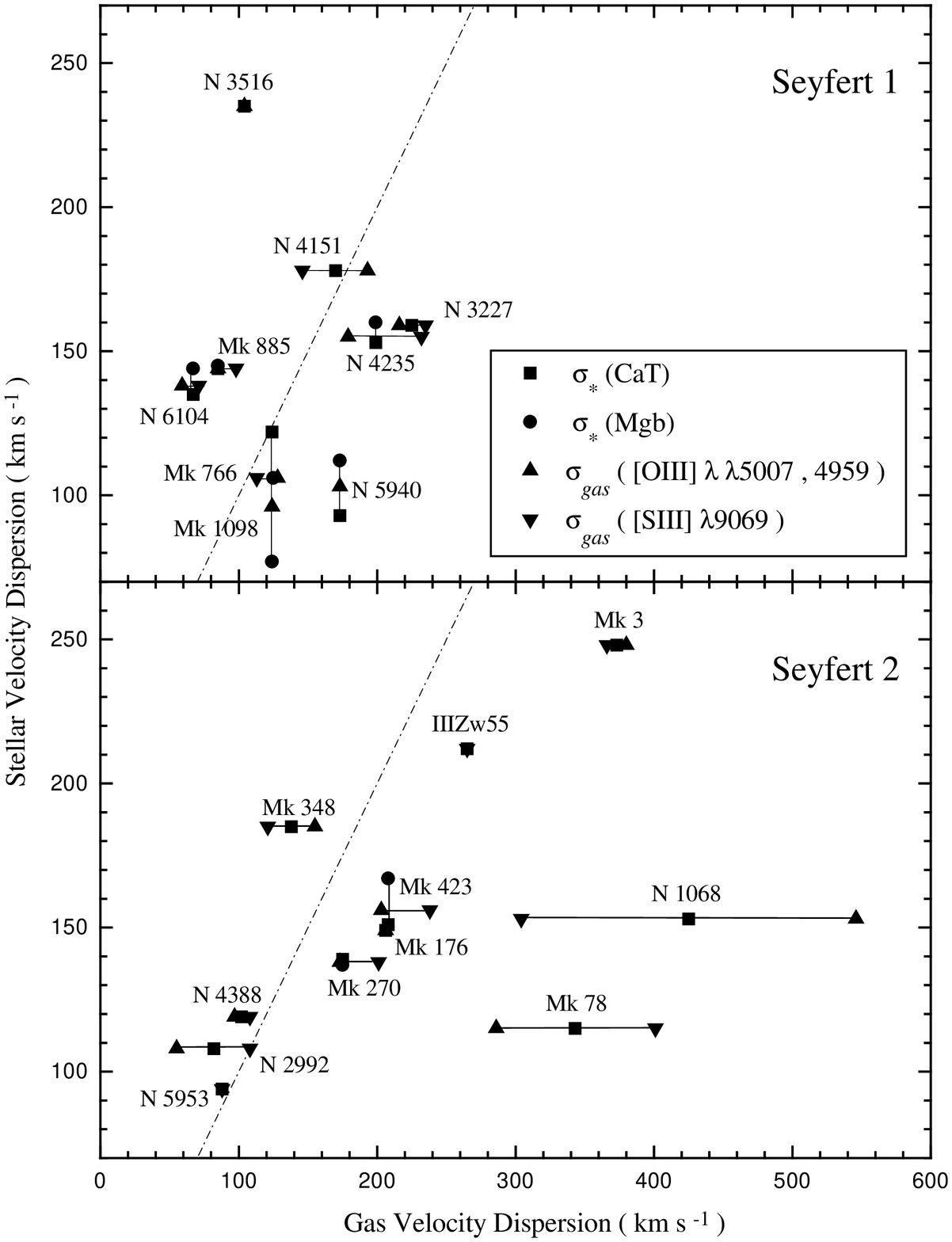,width=8cm,clip=}}
\caption{Stellar vs. gas velocity dispersion in Seyfert galaxies for
our new data together with the results from TDT90. 
There is no clear correlation between
the dispersion of the gas in the NLR and that of the stars in the nuclei 
of Seyferts although the majority
of the galaxies fall around the 1:1 line.}
\label{sy1sy2}
\end{figure}

\subsubsection{Young starbursts in the nuclei of Seyfert 1 galaxies}
\label{nuc_star}
All Seyfert 2 galaxies observed until now with good spatial resolution at UV wavelengths, show 
nuclear starbursts
(Heckman et~al. 1997; Gonz\'alez Delgado et~al. 1998). Thus, if unified models were correct, 
we should also see some starburst signatures in the spectra of  Seyfert 1 nuclei.
The presence of such starbursts in the nuclei of the group II galaxies,
could explain our observations, since the CaT is strong in the spectra of young red 
supergiants. 
On the other hand, only our luminous Seyfert 1 galaxies have the CaT relatively less diluted than
the Mgb. If similar starbursts were present in all Seyfert 1 nuclei, we should also see 
relatively strong CaT features in the nuclear spectra of the weaker Seyferts. 

This can be explained by the data on Seyfert 2 galaxies. Gonz\'alez Delgado et~al. \shortcite{Gon98} 
have found that the bolometric luminosities of the active nuclei and nuclear starbursts in Seyfert 
2 galaxies are positively related. 
If starbursts in
Seyfert 1 followed this same relation, the spectral features from young stars in the spectra 
of the weaker Seyfert 1 could be masked by the old stellar populations of the bulge. On the contrary,
in more active Seyfert, the starburst could be strong enough as to increase the EW of the CaT so that it
could be seen in spite of the strong continuum emitted by the AGN.

  \subsubsection{Blue indices diluted by Fe\,{\sevensize\it II} multiplets}
\label{Feii}
One of the common properties of all the galaxies in the
group II is that they are strong \hbox{Fe\,{\sc ii}} emitters. The lines of \hbox{Fe\,{\sc ii}},
(more precisely the bump formed by multiplets (42), (48) and (49)) 
can mask the Mgb, Fe$_{52}$ and Fe$_{53}$ stellar features, 
making these features  to appear diluted by the continuum emitted by the AGN. 
If this were true, the dilution of the three blue indices could be very different, since the emission of
\hbox{Fe\,{\sc ii}} is very irregular. 
This could explain why,
in some of our galaxies, specially in Mk 1098, the dilution of the Mgb is much greater than the dilution of the
Fe indices.  
On the other hand, 
the \hbox{Fe\,{\sc ii}} emission, in Seyfert galaxies, seems to be correlated with 
CaT emission \cite{Per88} i.e., the galaxies that are  strong \hbox{Fe\,{\sc ii}} emitters also show the CaT
in emission. This is what happens in Mk 766 and, perhaps, also in Mk 1098. In the other galaxies 
the CaT could also be in emission, 
at levels too low to be detected but high enough to dilute the EW of the absorption lines.

  \subsubsection{Compact SN remnants and evolutionary effects}

The hypothesis discussed in sections~\ref{nuc_star} and~\ref{Feii}
may actually be compatible with one another, since part of the \hbox{Fe\,{\sc ii}} 
emission
can be produced by compact supernova remnants (cSNRs) like  SN1987F 
(Filippenko 1989; Wegner \& Swanson 1996), SN1988Z (Stathakis \& Sadler 1991; Turatto et~al. 1993) or
SN1997ab (Hagen, Engels \& Reimers 1997; Salamanca et~al. 1998). 
These objects are characterized by showing broad permitted emission
lines of hydrogen, helium, \hbox{Fe\,{\sc ii}}, IR CaT, superposed on a nearly featureless continuum. 
Their optical spectra are so similar to those of QSOs and type 1 Seyferts, that they have 
been called ``Seyfert 1 imposters'' by Filippenko. 
Several explanations have been proposed for the intriguing properties of cSNRs. The most
promising are those based on the interaction of the SN ejecta with a dense circumstelar medium
($n \sim 10^7 \mbox{cm}^{-3}$) (Terlevich et~al. 1992a). 
In the starburst model for AGN of Terlevich et~al. 1992a,b
(see also the recent review of Cid Fernandes 1997), 
the broad permitted emission lines
of type 1 Seyfert nuclei are generated by strongly radiative cSNRs, in compact nuclear starbursts.
In this model, the differences between types 1 and 2 Seyferts are not only due to obscuration 
effects but also to the evolution of the central starburst, i.e. an AGN can only become of type 1 when the
nuclear starburst has reached  a phase at which their most massive stars begin to explode as type II SN
\cite{TMM87}. This happens when the burst has an age $>$8 Myr. 

The starburts discovered by Heckman et~al. \shortcite{Heck97} and Gonz\'alez 
Delgado et~al. \shortcite{Gon98} in four Seyfert 2 galaxies have ages between 
3 and 6 Myr. When these massive starbursts reach an age of 8 Myr, they should  
yield a very high SNe rate.
If, as expected from  numerical simulations, some of these SN
become  cSNRs, then these galaxies will, with a high probability, be 
classified as  type 1 AGN.

\begin{figure}
\centerline{
\psfig{figure=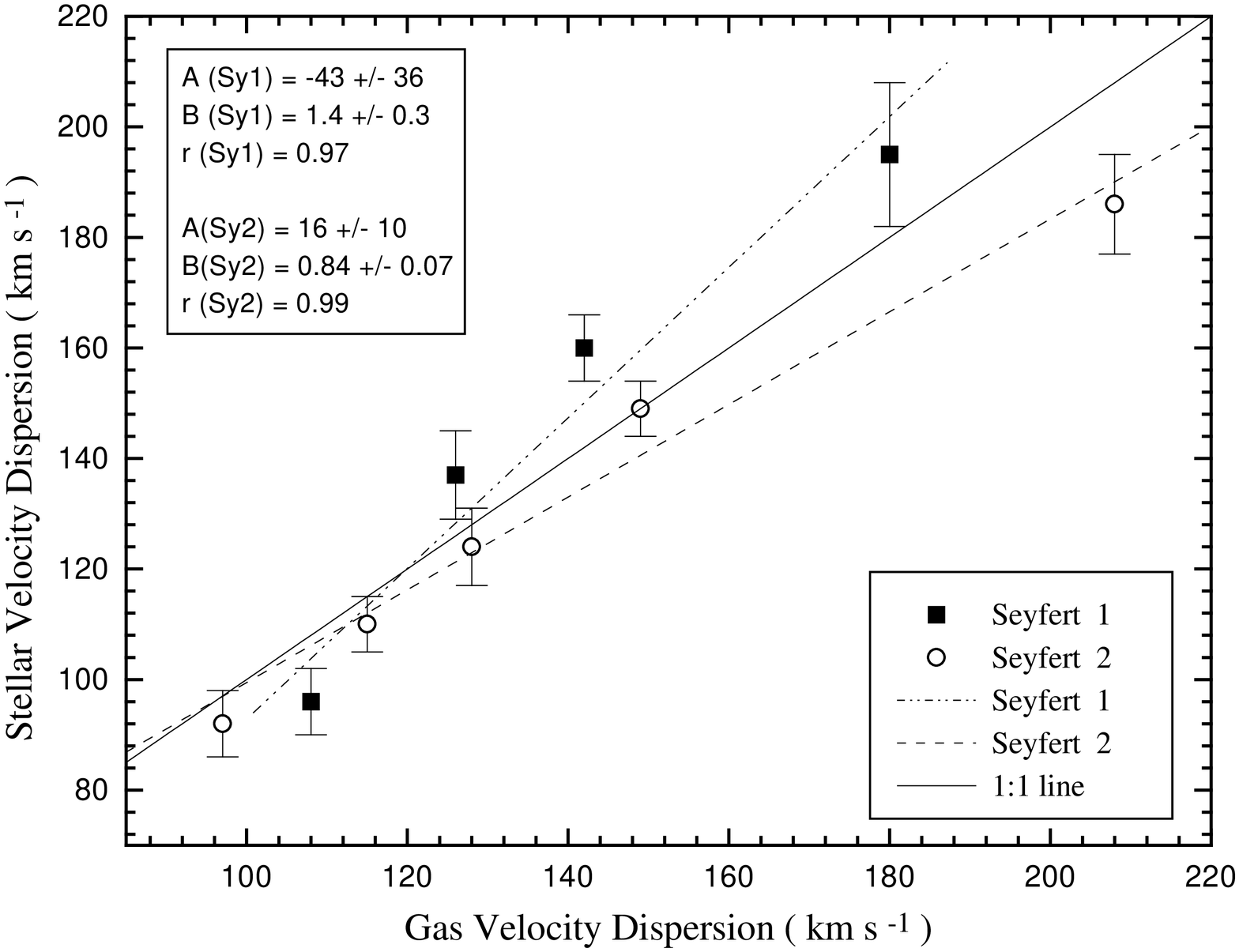,width=8cm,clip=}}
\caption{Stellar vs. gas velocity dispersions for the Seyfert galaxies of TDT90,  
Nelson \& Whittle (1995) and this work.
Seyferts 1 and 2 have been binned in stellar velocity dispersion and the 
medians of the two magnitudes for each bin are plotted.
They are found to be clearly correlated.}
\label{medians}
\end{figure}

 \subsection{Gas and stellar kinematics in Seyfert galaxies}

Gas in the NLR of Seyfert galaxies can be accelerated by the gravitational field of the galaxy or by 
violent processes, like shock waves or tidal forces.
This question has been analyzed in several papers
(Wilson \& Heckman 1985; TDT90; Nelson \& Whittle 1996), and all of them have arrived to the
following conclusions:

\begin{enumerate}
\item In the majority of Seyfert galaxies the motion of the gas in the NLR
seemed to be controlled by the gravitational field 
of the bulge, since the gas and stars show similar velocity dispersions.
\item In some Seyfert 2 galaxies the velocity dispersion of the gas is very 
high, rising to more than 400 km s$^{-1}$. In these galaxies extra broadening 
mechanisms must exist. Such mechanisms could be shock-waves caused by jets 
emitted by the active nucleus or gravitational perturbations caused by tidal
interactions with other galaxies. 
\end{enumerate} 

Fig.~\ref{sy1sy2} shows the stellar velocity dispersion vs. the width of the emission lines for our new data, 
together with those by TDT90. It can be 
seen in the figure that the majority
of the galaxies groups around the slope 1 line. 
A least square fit gives a correlation coefficient of 0.9 for Seyfert 2 galaxies (after eliminating
the points corresponding to NGC~1068 and Mk~78) and only 0.13 for Seyfert 1. The slopes of the lines fitted
are 0.48$\pm$0.09 for Seyfert 2 and 0.11$\pm$0.32 for Seyfert 1.
Thus, we cannot assert that the stellar and gas velocity dispersions are correlated in Seyfert galaxies.

The galaxies with emission lines much wider than
their stellar absorption lines
are all Seyfert 2. In these last galaxies, we also observe big differences between
the widths of the lines of \hbox{[O\,{\sc iii}]} and \hbox{[S\,{\sc iii}]} 
$\lambda$9069. This points to the existence of several gas clouds in the NLR 
of these galaxies, each one
with different ionization states and moving in different ways. In fact, 
these clouds have been resolved in NGC 1068
(e.g. Dietrich \& Wagner 1998; Kraemer et~al. 1998).

To improve the statistics, we added to our data that from Nelson \& Whittle
\shortcite{NW95}. If we represent the width of the \hbox{[O\,{\sc iii}]} $\lambda$5007 
emission line  vs. the stellar
velocity dispersion, we would see again a cloud of points around the slope 1 line, but not 
a clear correlation between both magnitudes.
Now, the correlation coefficient is 0.49 for Seyfert 2 and 0.44 for Seyfert 1. The slopes of the
lines fitted are 0.31$\pm$0.09 for Seyfert 2 and 0.4$\pm$0.2 for Seyfert 1.
To find out if the clustering of the data around 
the slope 1 line really means that the gas motions in Seyfert galaxies
are related to the stellar mass of their bulges, we have binned the Seyferts 1 and 2 
separatelly in stellar velocity dispersion
in such a way that there were  always more than 4 galaxies in each group. 
This results in 5 groups of type 2 Seyferts and 4 of type 1  
The median of the stellar velocity dispersions of the galaxies in each group
vs. the corresponding median of their emission line widths can be seen in  
Fig.~\ref{medians}. 
The correlation found between both medians is really surprising. 
A possible explanation is that the main mechanism that controls the gas motions in the
NLR is the gravitational field of the bulge; nevertheless, several other mechanisms might also influence 
the gas kinematics, therefore the correlation between the gas and stellar velocity dispersion is not
evident at first sight.
Nelson \& Whittle \shortcite{NW96}
arrived to this same conclusion by using a more elaborate method of analysis.

\begin{figure}
\centerline{
\psfig{figure=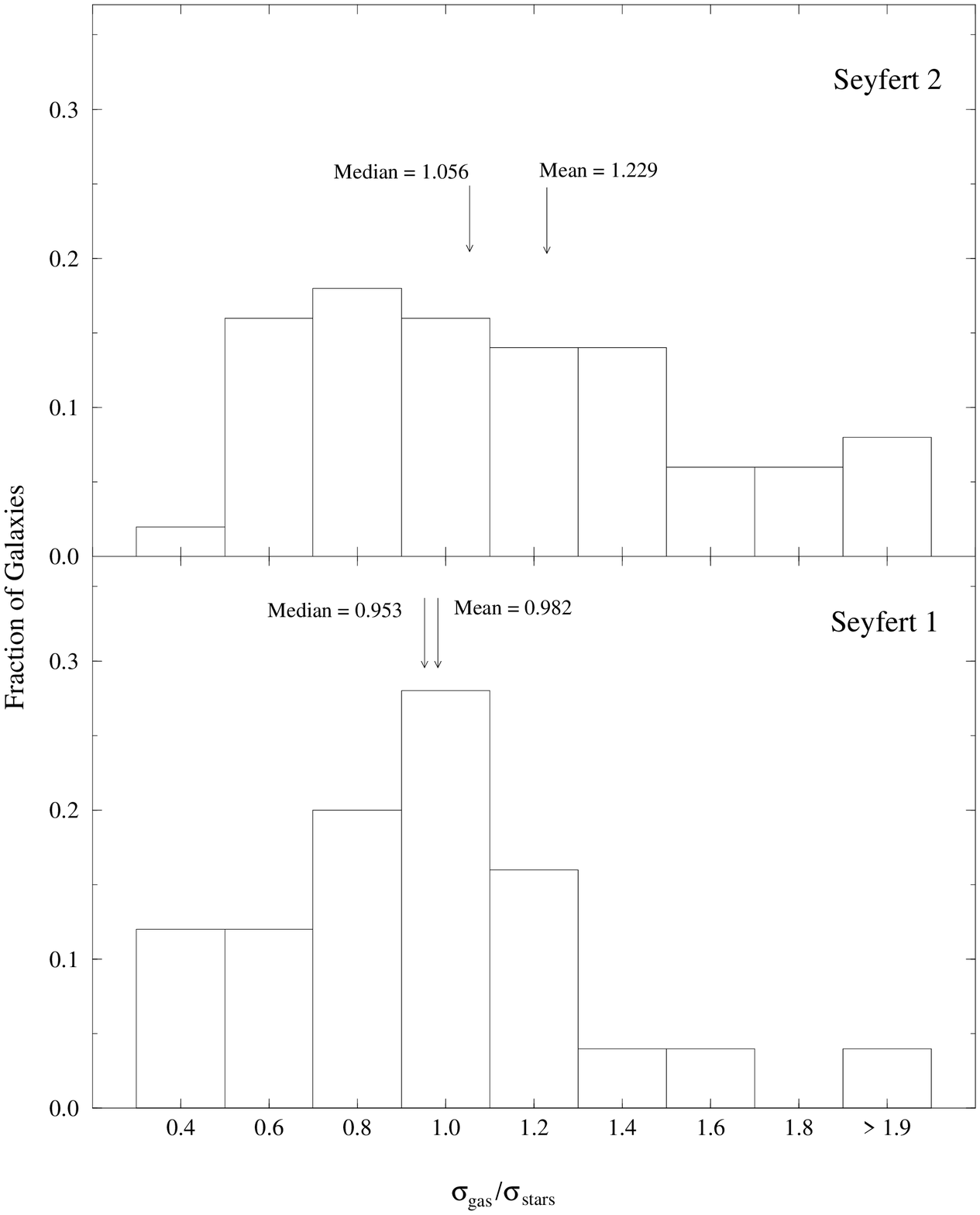,width=8cm,clip=}}
\caption{Ratio between the gas and stars velocity dispersion in Seyfert galaxies.}
\label{barras}
\end{figure}

Shock waves produced by jets (e.g. Falcke, Wilson \& Simpson 1998; Axon et~al. 1998)
and tidal effects caused by close encounters with other
galaxies can disturb and accelerate the gas in the NLR. This could explain why many Seyfert galaxies have
emission lines wider than their stellar absorption lines. 
However, it is more difficult to explain why in many other Seyfert galaxies
the emission lines are narrower than the stellar features. The reason could be that, in some galaxies, 
the gas has settled in a cold rotating disk. 
This become more intriguing if we realize
that Seyfert 1 and 2 show different behaviours in Fig.~\ref{medians}. 
The medians of the gas velocity in type 2 Seyferts are
always higher than the ones of the stellar velocity dispersions, while the contrary happens in Seyfert 1.
Fig.~\ref{barras} represents the ratio
between the velocity dispersion of gas and stars in Seyferts,
and we can see that, whereas there 
is a population of type 2 Seyferts with gas velocities much higher than the velocities of their stars, 
in Seyfert 1
there are signs of the existence of another population with inverse kinematical properties. This means that
the differences between the kinematical behaviour of Seyfert 1 and 2  are not likely due to 
systematic differences between all the galaxies of both groups, but to the existence of these two
populations. 

There are several possibilities for explaining the different
kinematical behaviours of Seyfert 1 and 2 galaxies.
If shock-waves produced by jets were the main accelerating mechanism of the gas in Seyferts (aside of the 
gravitational field),
the lack of Seyfert 1 galaxies with emission lines much broader than their absorption features, could be
explained in the context of unified models. According to this model the active nucleus in a Seyfert galaxy is
surrounded by a dense molecular torus. Therefore, if the AGN produced radio-jets, they would be emitted 
in a direction perpendicular to the
torus plane. The gas in the NLR would also be accelerated in this same direction. Therefore, 
if the line of sight to the
AGN is perpendicular to the torus plane ( as, according to unified models, is 
the case for Seyfert 1)
we should not see a broadening in the emission lines of the galaxy, but a spectral shift between the stellar
and gas emission features.

On the other hand,
if the nuclear activity, in at least some Seyfert galaxies, is produced by gravitational encounters with 
close companions (e.g. Keel 1996; De Robertis, Yee \& Hayhoe 1998),
the younger of these Seyferts should also be those with broader emission lines. 
Tidal interactions can disturb 
and accelerate the gas in the nuclei of the involved galaxies, but, immediately after the encounter, 
the gas in both galaxies begins to settle again so that in a few Myr the kinematical signs of the interaction
will have disappeared. Therefore, the observed differences between the gas velocity dispersions of Seyfert 1 and 2
could be explained if nuclear activity were older in Seyfert 1 than in Seyfert 2. 

We could also appeal to morphological differences between Seyfert 1 and 2 to explain their kinematical behaviour.
Seyfert 1 are frequently found in earlier galaxies than Seyfert 2 \cite{MGT98}. The gas and stellar velocity 
dispersions in starbursts 
found in late type spirals  are of the same order, whereas in early type 
starburst galaxies the stellar velocity dispersion seems to be higher than that of the gas \cite{Vega98}.
This difference between the gas and stellar kinematics in starburst galaxies could extend to Seyfert galaxies.
To test this last possibility, we have compared the differences between the velocity dispersions of the gas and
stars in Seyfert galaxies vs. their  Hubble type and their stellar velocity dispersion (which is a measurement
of the bulge mass)
but we have not found
any clear correlation between these magnitudes. Therefore, we must conclude that it is not likely that
the kinematical differences 
between Seyfert 1 and 2 be due just to morphological differences between both types of galaxies.



\section{Conclusions}

We have studied the CaT and the Mgb stellar absorption indices and 
kinematics of the nuclei of a sample of ten active galaxies, 
centering our analysis on the study of the properties of Seyfert 1 
galaxies.

In spite of problems related to contamination by emission lines, 
we find that the IR CaII triplet stellar indices of Seyfert 1 nuclei 
with EW(H$\beta_{broad}$) $\la$ 45 \AA, 
is stronger than what the standard model would predict from the observed 
strength of the Mgb index. 

This result is naturally explained by the presence, in the nuclei 
of these type 1 Seyferts of young stellar clusters whose luminosity 
is somehow related to the luminosity of the active nucleus itself.
This conclusion is weaker for those nuclei with strong \hbox{Fe\,{\sc ii}} 
emission affecting the measurement of the Mgb  index.

Our measurements of the velocity dispersions in Seyfert galaxies, support 
previous conclusions by TDT90 and Nelson \& Whittle \shortcite{NW96} that the 
main factor that controls the gas motions in 
the NLR of Seyfert galaxies is the mass of the bulge. We also find that 
other factors may also be important, since the correlation between the gas 
and stellar velocity dispersions shows a large scatter. Among the factors that 
may broaden the emission lines in the NLR, induced motions by shocks are
the most likely. 

Some differences between the kinematics of Seyfert 1 and Seyfert 2 have been 
found. Seyfert galaxies with 
gas velocity dispersions much larger than that of the stars
are preferentially found among type 2. Also
we found some Seyfert 1 nuclei with emission lines narrower than the stellar
absorption features.
Although we have outlined some hypotheses, like orientation an evolutionary 
effects,
a satisfactory explanation for this second aspect still needs to be found.



\section*{Acknowledgments}

We would like to thank Itziar Aretxaga, Javier Gorgas, Enrique P\'erez, Javier 
Palacios and Juan Carlos Vega for many helpful suggestions. We also thank the 
staffs at the WHT and the 
ING for their assistance and support. E.T.~acknowledges an IBERDROLA Visiting 
Professorship to UAM. L.J.B. thanks the hospitality of INAOE and of the 
Guillermo Haro  Programme for Advanced Astrophysics during the finalization 
of this paper.
This research has made use of NASA's Astrophysics Data System (ADS) 
and of the NASA/IPAC Extragalactic Database (NED).




\label{lastpage}

\begin{thebibliography}{}


\bibitem[\protect\citename{Abell et al.\ }1978]{AEJ78}
Abell~G.~O., Eastmond~T.~J., Jenner~C., 1978,
ApJ, 221, L1

\bibitem[\protect\citename{Antonucci }1993]{Ant93}
Antonucci~R.~R., 1993, 
ARA\&A, 31, 473

\bibitem[\protect\citename{Axon }1998]{Axon98}
Axon~D.~J., Marconi~A., Capetti~A., Maccetto~F.~D., Schreier~E., Robinson~A., 1998,
ApJ, 496, 75

\bibitem[\protect\citename{Bica et al.\ }1991]{Bic91}
Bica~E., Pastoriza~M.~G., Maia~M., da Silva~L.~A.~L., Dottori~H., 1991,
AJ, 102, 1702

\bibitem[\protect\citename{Brotherton et al.\ }1999]{BBS99}
Brotherton~M.~S., van Breugel~W., Stanford~S.~A., Smith~R.~J., Boyle~B.~J.,
Miller~L., Shanks~T., Croom~S.~M., Filippenko~A.~V., 1999, ApJ in press, 
astro-ph/9906052

\bibitem[\protect\citename{Cardiel, Gorgas \& Arag\'on--Salamanca }1998]{CGA98}
Cardiel~N., Gorgas~J., Arag\'on--Salamanca~A., 1998,
MNRAS, 298, 977

\bibitem[\protect\citename{Cid Fernandes }1997]{Cid97}
Cid Fernandes~R., 1997,
RMxAC, 6, 201

\bibitem[\protect\citename{Cid Fernandes \& Terlevich }1992]{CT92}
Cid Fernandes~R., Terlevich~R., 1992,
in Filippenko, A.~V., ed., ASP Conf.\ Ser.\ Vol.\ 31,
Relationships between Active Galactic Nuclei and Starburst Galaxies.
Astron.\ Soc.\ Pac., San Francisco, p.\ 241

\bibitem[\protect\citename{Cid Fernandes \& Terlevich }1993]{CT93}
Cid Fernandes~R., Terlevich~R., 1993,
in Beckman J., Netzer H., Colina L., eds, The Nearest Active Galaxies.
Ap\&SpSci 205, 91

\bibitem[\protect\citename{Cid Fernandes \& Terlevich }1995]{CT95}
Cid Fernandes~R., Terlevich~R., 1995,
MNRAS, 272, 423

\bibitem[\protect\citename{Colbert et al.\ }1996a]{Col96a}
Colbert~E.~J.~M., Baum~S.~A., Gallimore~J.~F., O'Dea~C.~P., Lehnert~M.~D., 
Tsvetanov~Z.~I., Mulchaey~J.~S., Caganoff~S., 1996a,
ApJS, 105, 75

\bibitem[\protect\citename{Colbert et al.\ }1996b]{Col96b}
Colbert~E.~J.~M., Baum~S.~A., Gallimore~J.~F., O'Dea~C.~P., Christensen~J.~A.,
1996b,
ApJ, 467, 551

\bibitem[\protect\citename{Colina et al.\ }1997]{Col98}
Colina~L., Garc\'{\i}a--Vargas~M.~L., Mas--Hesse~J.~M, Alberdi~A., Krabbe~A.,1997
ApJ, 484, L41

\bibitem[\protect\citename{Dahari \& De Robertis }1989]{DR89}
Dahari~D., De Robertis~M.~M., 1989,
ApJ, 252, 102

\bibitem[\protect\citename{De Robertis et al.\ }1998]{RYH98}
De Robertis~M.~M., Yee~H.~K.~C., Hayhoe~K., 1998,
ApJ, 496, 93

\bibitem[\protect\citename{de Vaucouleurs et al.\ }1991]{RC3}
de Vaucouleurs~G., de Vaucouleurs~A., Corwin~H.~G.~Jr., Buta~R.~J., Paturel~G., Fouqu\'e~P., 1991,
The Third Reference Catalogue of Bright Galaxies. Springer-Verlag, New York

\bibitem[\protect\citename{D\'{\i}az et al.\ }1989]{DTT89}
D\'{\i}az~A.~I., Terlevich~E., Terlevich~R., 1989,
MNRAS, 239, 325

\bibitem[\protect\citename{Dietrich \& Wagner }1998]{DW98}
Dietrich~M., Wagner~S.~J., 1998,
A\&A, 338, 405

\bibitem[\protect\citename{Falcke et al.\ }1998]{FWS98}
Falcke~H., Wilson~A.~S., Simpson~C, 1998,
ApJ, 502, 199

\bibitem[\protect\citename{Filippenko }1989]{Fil89}
Filippenko~A.~V., 1989,
AJ, 97, 726

\bibitem[\protect\citename{Fisher et al.\ }1996]{FFI96}
Fisher~D., Franx~M., Illingworth~G., 1996,
ApJ, 459, 110

\bibitem[\protect\citename{Garc\'{\i}a--Vargas et al.\ }1998]{GMB98}
Garc\'{\i}a--Vargas~M.~L., Moll\'a~M., Bressan~A., 1998,
A\&AS, 130, 513

\bibitem[\protect\citename{Gonz\'alez Delgado \& P\'erez }1993]{GP93}
Gonz\'alez Delgado~R.~M., P\'erez~E., 1993,
Ap\&SS, 205, 127

\bibitem[\protect\citename{Gonz\'alez Delgado \& P\'erez }1996]{GP96}
Gonz\'alez Delgado~R.~M., P\'erez~E., 1996,
MNRAS, 278, 737

\bibitem[\protect\citename{Gonz\'alez Delgado et al.\ }1997]{Gon97}
Gonz\'alez Delgado~R.~M., P\'erez~E., Tadhunter~C., V\'{\i}lchez~J.~M., 
Rodr\'{\i}guez--Espinosa~J.~M., 1997,
ApJS, 108, 155

\bibitem[\protect\citename{Gonz\'alez Delgado et al.\ }1998]{Gon98}
Gonz\'alez Delgado~R.~M., Heckman~T.~M., Leitherer~C., Meurer~G., 
Krolik~J., Wilson~A.~S., Kinney~A., Koratkar~A., 1998, 
ApJ, 505, 174

\bibitem[\protect\citename{Goodrich }1989]{Good89}
Goodrich~R.~W., 1989,
ApJ, 342, 224

\bibitem[\protect\citename{Gorgas et al.\ }1993]{Gor93}
Gorgas~J., Faber~S.~M.,Burstein~D., Gonz\'alez~J., Courteau~S., Prosser~C., 1993,
ApJS, 86, 153

\bibitem[\protect\citename{Goudfrooij \& Emsellem }1996]{GE96}
Goudfrooij~P., Emsellem~E., 1996,
A\&A, 306, L45

\bibitem[\protect\citename{Hagen, Engels \& Reimers }1997]{HER97}
Hagen~H.~J., Engels~D., Reimers~D., 1997,
A\&A, 324, L29

\bibitem[\protect\citename{Haniff et al.\ }1988]{HWW88}
Haniff~C.~A., Wilson~A.~S., Ward~M.~J., 1988,
ApJ, 334, 104

\bibitem[\protect\citename{Heckman et al.\ }1997]{Heck97}
Heckman~T.~M., Gonz\'alez--Delgado~R.~M., Leitherer~C., Meurer G.~R., 
Krolik~J., Wilson~A.~S., Koratkar~A., Kinney~A., 1997,
ApJ, 482, 114

\bibitem[\protect\citename{Ho, Filippenko \& Sargent }1997]{HFS97}
Ho~L.~C., Filippenko~A.~V., Sargent~W.~L.~W., 1997,
ApJ, 487, 568

\bibitem[\protect\citename{Idiart et al.\ }1996]{IFC96}
Idiart~T.~P., de Freitas Pacheco~J.~A., Costa~R.~D.~D., 1996,
AJ, 112, 2541

\bibitem[\protect\citename{Jones et al.\ }1984]{JAJ84}
Jones~J.~E., Alloin~D.~M., Jones~B.~J.~T., 1984,
ApJ, 283, 457

\bibitem[\protect\citename{Keel }1996]{Keel96}
Keel~W.~C., 1996,
AJ, 111, 696

\bibitem[\protect\citename{Kraemer et al.\ }1998]{Kra98}
Kraemer~S.~B., Ruiz~J.~R., Crenshaw~D.~M., 1998,
ApJ, 508, 232

\bibitem[\protect\citename{Kukula et al.\ }1995]{Kuk95}
Kukula~M.~J., Pedlar~A., Baum~S.~A.,  O'Dea~C.~P., 1995,
MNRAS, 276, 1262

\bibitem[\protect\citename{Maiolino et al.\ }1995]{Mai95}
Maiolino~R., Ruiz~M., Rieke~G.~H., Keller~L.~D., 1995,
ApJ, 446, 561

\bibitem[\protect\citename{Malkan et al.\ }1998]{MGT98}
Malkan~M.~A., Gorjian~V., Tam~R., 1998,
ApJS, 117, 25

\bibitem[\protect\citename{Maoz et al.\ }1998]{Maoz98}
Maoz~D., Koratkar~A., Shields~J.~C., Ho~L.~C., Filippenko~A.~V. 
Sternberg~A., 1998,
AJ, 116, 55

\bibitem[\protect\citename{Molendi \& Maccacaro }1994]{MM94}
Molendi~S., Maccacaro~T., 1994,
A\&A, 291, 420

\bibitem[\protect\citename{Molendi et al.\ }1993]{MMS93}
Molendi~S., Maccacaro~T., Schaeidt~S., 1993,
A\&A, 271, 18

\bibitem[\protect\citename{Morris \& Ward }1988]{MW88}
Morris~S.~L., Ward~M.~J., 1988,
MNRAS, 230, 639

\bibitem[\protect\citename{Mulchaey et al.\ }1996]{MWT96}
Mulchaey~J.~S., Wilson~A.~S., Tsvetanov~Z., 1996,
ApJS, 102, 309

\bibitem[\protect\citename{Nandra et al.\ }1997]{Nan97}
Nandra~K., George~I.~M., Mushotzky~R.~F., Turner~T.~J., Yaqoob~T., 1997,
ApJ, 477, 602

\bibitem[\protect\citename{Nelson \& Whittle }1995]{NW95}
Nelson~C.~H., Whittle~M., 1995,
ApJS, 99, 67

\bibitem[\protect\citename{Nelson \& Whittle }1996]{NW96}
Nelson~C.~H., Whittle~M., 1996,
ApJ, 465, 96

\bibitem[\protect\citename{Netzer }1990]{Net90}
Netzer~H., 1990,
in Courvoisier~T.~J.~L., Mayor~M., eds, 
Active Galactic Nuclei. Springer--Verlag, Berlin, p. 57

\bibitem[\protect\citename{Nordgren et al.\ }1995]{Nor95}
Nordgren~T.~E., Helou~G., Chengalur~J.~N, Terzian~Y., Khachikian~E.~D., 1995,
ApJS, 99, 461

\bibitem[\protect\citename{Oliva et al.\ }1995]{Oli95}
Oliva~E., Origlia~L., Kotilainen~J.~K., Moorwood~A.~F.~M., 1995,
A\&A, 301, 55

\bibitem[\protect\citename{Osterbrock }1977]{Ost77}
Osterbrock~D.~E., 1977,
ApJ, 215, 733

\bibitem[\protect\citename{Osterbrock }1981]{Ost81}
Osterbrock~D.~E., 1981,
ApJ, 249, 462

\bibitem[\protect\citename{Osterbrock \& Dahari }1983]{OD83}
Osterbrock~D.~E., Dahari~O., 1983,
ApJ, 273, 478

\bibitem[\protect\citename{Osterbrock \& Pogge }1985]{OP85}
Osterbrock~D.~E., Pogge~R.~W., 1985,
ApJ, 297, 166

\bibitem[\protect\citename{Palacios et al.\ }1997]{Pal97}
Palacios~J., Garc\'{\i}a--Vargas~M.~L., D\'{\i}az~A.~I., Terlevich~R., Terlevich~E., 
1997, A\&A,  323, 749

\bibitem[\protect\citename{Persson }1988]{Per88}
Persson~S.~E., 1988,
ApJ, 330, 751

\bibitem[\protect\citename{Pogge }1989]{Pog89}
Pogge~R.~W., 1989,
ApJ, 345, 730

\bibitem[\protect\citename{Rudy et al.\ }1985]{RCP85}
Rudy~R.~J., Cohen~R.~D., Puetter~R.~C., 1985,
ApJ, 288, L29

\bibitem[\protect\citename{Salamanca et al.\ }1998]{Sal98}
Salamanca~I., Cid Fernandes~R., Tenorio--Tagle~G., Telles~E., Terlevich~R., 
Mu\~noz--Tu\~n\'on, 1998,
MNRAS, 300, L17

\bibitem[\protect\citename{Schmitt et al.\ }1999]{SSC98}
Schmitt~H.~R., Storchi--Bergmann~T., Cid Fernandes~R., 1999,
MNRAS, 303, 173

\bibitem[\protect\citename{Serote--Roos et al.\ }1996]{Ser96}
Serote--Roos~M., Boisson~C., Joly~M., Ward~M.~J., 1996,
MNRAS, 278, 897

\bibitem[\protect\citename{Simien \& de Vaucouleurs }1986]{SV86}
Simien~F., de Vaucouleurs~G., 1986,
ApJ, 302, 564

\bibitem[\protect\citename{Stathakis \& Sadler }1991]{SS91}
Stathakis~R.~A., Sadler~E.~M., 1991,
MNRAS, 250, 786

\bibitem[\protect\citename{Terlevich et al.\ }1990]{TDT90}
Terlevich~E., D\'{\i}az~A.~I., Terlevich~R., 1990, 
MNRAS, 242, 271

\bibitem[\protect\citename{Terlevich, Melnick \& Moles }1987]{TMM87}
Terlevich~R., Melnick~J., Moles~M., 1987,
in Khachikian~E., Fricke~K., Melnick~J., eds,
Observational Evidence for Activity in Galaxies.
Reidel, Dordrecht, p.~499

\bibitem[\protect\citename{Terlevich et al.\ }1992a]{Ter92a}
Terlevich~R., Tenorio--Tagle~G., Franco~J., Melnick~J., 1992a,
MNRAS, 255, 713

\bibitem[\protect\citename{Terlevich et al.\ }1992b]{Ter92b}
Terlevich~R., Tenorio--Tagle~G., R\'ozyczka~M., Franco~J., Melnick~J., 1992b,
MNRAS, 272, 198

\bibitem[\protect\citename{Tory \& Davis }1979]{TD79}
Tonry~J., Davis~M., 1979,
AJ, 84, 1511

\bibitem[\protect\citename{Turatto et al.\ }1993]{Tur93}
Turatto~M., Cappellaro~E., Danziger~I.~J., Benetti~S., Gouiffes~C., Della Valle~M., 1993,
MNRAS, 262, 128

\bibitem[\protect\citename{Ulvestad }1986]{Ulv86}
Ulvestad~J.~S., 1986,
ApJ, 310, 136

\bibitem[\protect\citename{Ulvestad \& Wilson }1984a]{UW84a}
Ulvestad~J.~S., Wilson~A.~S., 1984a,
ApJ, 278, 544

\bibitem[\protect\citename{Ulvestad \& Wilson }1984b]{UW84b}
Ulvestad~J.~S., Wilson~A.~S., 1984b,
ApJ, 285, 439

\bibitem[\protect\citename{Vega et al.\ }1998]{Vega98}
Vega Beltr\'an~J.~C., Pignatelli~E., Zeilinger~W., Bertola~F,
Beckman~J.E., Pizzella~A., Corsini~E., 1998,
in Merritt~D.~R., Valluri~M., Sellwod~J.~A., eds, 
ASP Conf.\ Ser., 
Galaxy Dynamics. Astron.\ Soc.\ Pac., San Francisco, in press, astro-ph/9811235 

\bibitem[\protect\citename{Veilleux }1991]{Vei91}
Veilleux~S., 1991,
ApJ, 369, 331

\bibitem[\protect\citename{Wegner \& Swanson }1996]{WS96}
Wegner~G., Swanson~S.~R., 1996,
MNRAS, 278, 22

\bibitem[\protect\citename{Wilson \& Heckman }1985]{WH85}
Wilson~A.~S., Heckman~T.~M., 1985,
in Miller~J.~S., ed., Astrophysics of Active Galaxies and Quasi--Stellar Objects.
University Science Books, Mill Valley, CA, p. 39

\bibitem[\protect\citename{Xanthopoulos \& De Robertis }1991]{XR91}
Xanthopoulos~E., de~Robertis~M.~M., 1991,
AJ, 102, 1980

\end{thebibliography}
\end{document}